\documentclass[%
 reprint,
superscriptaddress,
 amsmath,amssymb,
 aps,
]{revtex4-2}

\usepackage{graphicx}
\usepackage{bm}

\usepackage{upgreek}
\usepackage{array}

\begin{document}

\title{Diagonalizing an optical coherence matrix via on-chip Stokes tomography}

\author{Amin Hashemi}
\affiliation{CREOL, The College of Optics \& Photonics, University of Central Florida, Orlando, FL 32816, USA}
\affiliation{These authors contributed equally to this work}

\author{Abbas Shiri}
\affiliation{CREOL, The College of Optics \& Photonics, University of Central Florida, Orlando, FL 32816, USA}
\affiliation{These authors contributed equally to this work}

\author{Bahaa E. A. Saleh}
\affiliation{CREOL, The College of Optics \& Photonics, University of Central Florida, Orlando, FL 32816, USA}

\author{Andrea Blanco-Redondo}
\affiliation{CREOL, The College of Optics \& Photonics, University of Central Florida, Orlando, FL 32816, USA}

\author{Ayman F. Abouraddy}
\affiliation{CREOL, The College of Optics \& Photonics, University of Central Florida, Orlando, FL 32816, USA}
\affiliation{raddy@creol.ucf.edu}

\begin{abstract}
Structured coherence -- partially coherent light spanned by a finite number of modes -- is emerging as a powerful tool in optical communications, computation, cryptography, and spectroscopy. Key to these prospects is the recent development of on-chip processing of structured coherence, in which large meshes of interferometers implement unitary and non-unitary transformations on the Hermitian coherence matrix representing multimode partially coherent light. Two related critical tasks for the applications of structured coherence are the reconstruction of an unknown coherence matrix and its diagonalization. Stokes tomography has been utilized in reconstructing the coherence matrix, whereas variational processing has been employed in its diagonalization. We show here that Stokes tomography can also be exploited in the on-chip diagonalization of an unknown coherence matrix, which we verify for two-mode and four-mode structured coherence in an integrated hexagonal mesh of Mach-Zehnder interferometers. This photonic circuit implements a predetermined sequence of configurations to estimate the generalized Stokes parameters, which -- in a final step -- inform a reconfiguration of the photonic circuit that diagonalizes the coherence matrix. The field is thus left in a coherent-mode representation comprising uncorrelated, orthogonal modes whose weights correspond to the eigenvalues of the original coherence matrix. We verify the coherent-mode representation of the diagonalized field by measuring the modal weights and comparing them to the eigenvalues of the original coherence matrix, and -- independently -- via on-chip interferometry to confirm that the modes are mutually uncorrelated. Moreover, the integrated photonic circuit can be configured to provide the original field alongside its diagonalized counterpart at the circuit output. We verify the diagonalization procedure for coherence matrices of different coherence rank, entropy, and structure. Finally, we dispel the common notion that $\mathcal{O}(N^{2})$ steps are required for reconstructing an $N\times N$ coherence matrix and show that only $\mathcal{O}(N)$ steps are needed. Stokes tomography is thus a versatile tool for on-chip characterization of structured coherence.
\end{abstract}


\maketitle

\section{Introduction}

The study of partially coherent light over the past two centuries has been restricted to freely propagating fields \cite{Born99Book,Wolf07Book,Gbur10PO,Korotkova20PO} (with few exceptions \cite{Saleh81AO,Saleh25book}). The past year has witnessed rapid progress in the \textit{on-chip} manipulation of multimode partially coherent light that is spanned by a finite modal basis, which we have denoted `structured coherence' \cite{Abouraddy26OPN,Hashemi26arxivTwoMode,Hashemi26arxiv4Modes,Hashemi26OL,Hashemi26LPR3modes,Abouraddy26AOP}. The transition from freely propagating fields to on-chip structured coherence \cite{roques2024Light,Miller25Optica,Mor26arxiv} can enable advanced studies of the dynamics of coherence entropy \cite{Okoro17Optica,Harling22OE,Harling23JO} and the coherence rank \cite{Harling24PRA,Harling24PRA2}, and can help exploit the `coherence advantage': scenarios in optical communications and signal processing in which partially coherent light outperforms coherent light \cite{Abouraddy26OPN,Abouraddy26AOP}. Recent examples of the coherence advantage making use of structured coherence include applications in optical computing \cite{Dong24Nature}, cryptography \cite{Peng21P,Liu23PRAppl,Liu25LPR}, spectroscopy \cite{Miller25Optica,Valdez26exp}, dense optical communications \cite{Nardi22OL}, and scattering-free optical communications \cite{Harling25APLP}.

A crucial task in the on-chip processing of structured coherence is the reconstruction of an unknown $N\times N$ Hermitian coherence matrix $\mathbf{G}$ supported by $N$~modes. Another task is the diagonalization of $\mathbf{G}$, whereupon $\mathbf{G}$ is converted via a unitary transformation $\hat{U}$ (`unitary' henceforth for brevity) into a `coherent-mode' representation $\mathbf{G}^{\mathrm{D}}$ \cite{Wolf86JOSAA}, where the transformed field is supported by orthogonal, mutually uncorrelated modes whose modal weights correspond to the eigenvalues of $\mathbf{G}$. These two tasks -- reconstructing and diagonalizing $\mathbf{G}$ -- are obviously related. Once $\mathbf{G}$ is reconstructed, it can be computationally diagonalized. Conversely, obtaining $\mathbf{G}^{\mathrm{D}}$ by physically diagonalizing $\mathbf{G}$ involves computationally reconstructing the unitary $\hat{U}$ from which $\mathbf{G}$ can be reconstructed. It is nevertheless useful to have access to both capabilities on chip in a single setting: reconstructing $\mathbf{G}$ \textit{and} physically providing the field in diagonalized form $\mathbf{G}^{\mathrm{D}}$. In our previous work we reconstructed $\mathbf{G}$ for $N=2$ \cite{Hashemi26arxivTwoMode,Hashemi26OL}, $N=3$ \cite{Hashemi26LPR3modes}, and $N=4$ \cite{Hashemi26arxiv4Modes} via Stokes tomography, in which the generalized modal Stokes parameters (SPs) are obtained as an intermediary to reconstructing $\mathbf{G}$. However, we did not physically diagonalize $\mathbf{G}$ on-chip to leave it in a coherent-mode representation. Recently, an alternate approach based on variational processing \cite{roques2024Light} was used to diagonalize the coherence matrix $\mathbf{G}$ \cite{Mor26arxiv}, but requires a significantly larger number of steps to accomplish this task than the number of steps required to reconstruct $\mathbf{G}$ via Stokes tomography.

Here we show that Stokes tomography enables both on-chip diagonalization of an unknown coherence matrix \textit{and} its reconstruction in the same platform. We make use of a photonic integrated circuit comprising a hexagonal mesh of Mach-Zehnder interferometers (MZIs) constructed out of \textit{tunable} phase shifters and \textit{fixed} symmetric mode couplers \cite{Bogaerts20Nature,Capmany20Book}. By tuning these phase shifters, the MZI mesh can implement an arbitrary $N\times N$ unitary on $N$~modes. We first cycle through a predetermined sequence of steps implemented to recover the modal SPs. In each step, a new unitary is implemented and the output modal weights are recorded. These measurements enable reconstructing $\mathbf{G}$ from which we calculate the diagonalizing unitary $\hat{V}$. In the final step, we reconfigure the on-chip unitary to realize $\hat{V}$ to diagoanlize the coherence matrix $\mathbf{G}\rightarrow\hat{V}\mathbf{G}\hat{V}^{\dagger}=\mathbf{G}^{\mathrm{D}}$. We carry out this procedure for two-mode fields ($N=2$) and four-mode fields ($N=4$). Crucially, we dispel the common notion that Stokes tomography requires $\mathcal{O}(N^{2})$ steps to reconstruct an $N\times N$ Hermitian coherence matrix \cite{Mor26arxiv}. Rather, only $\mathcal{O}(N)$ steps are needed to accomplish this task as long as all the output modal weights are recorded in each step. Specifically, $2N-1$ steps are needed to reconstruct $\mathbf{G}$ when $N$~is even, and $2N+1$ when $N$~is odd. Moreover, the on-chip MZI mesh allows for both $\mathbf{G}$ and its diagonalized counterpart $\mathbf{G}^{\mathrm{D}}$ to be simultaneously provided at the chip output for further processing. We confirm that the coherent-mode representation comprises mutually uncorrelated modes by verifying the lack of interference fringes when they are superposed in an on-chip interferometer -- although the modes are originally partially correlated before diagonalization. These results highlight the versatility of Stokes tomography as a general strategy for on-chip characterization of structured coherence.

\section{Diagonalization procedure}

\subsection{Overall strategy}

\begin{figure}[t!]
\centering
\includegraphics[width=8.8cm]{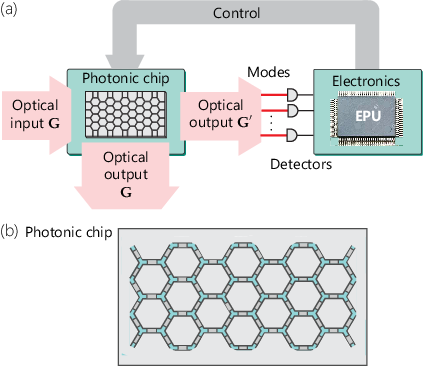}
\caption{(a) The input optical field (described by the coherence matrix $\mathbf{G}$) is coupled to an integrated photonic circuit that transforms the field with a programmable circuit. The structure of the implemented on-chip photonic circuit is determined by an electrical control signal and the coherence matrix associated with the resulting field is denoted $\mathbf{G}'$. The modal weights of the field, corresponding to the diagonal elements of $\mathbf{G}'$, are detected and stored in an electrical processing unit (EPU), and may influence future configurations of the photonic circuit through feedback. A copy of the input field $\mathbf{G}$ can also be made available for separate processing. (b) Structure of the hexagonal MZI mesh used in our experiments. The lines and gray rectangles are on-chip waveguides and MZI's, respectively.}
\label{fig:GeneralConcept}
\end{figure}

The conceptual scheme for on-chip diagonalization of a coherence matrix via Stokes tomography is depicted in Fig.~\ref{fig:GeneralConcept}(a). An optical field spanned by $N$ modes (e.g., the field in $N$~single-mode fibers) -- whose coherence matrix is denoted $\mathbf{G}$ -- is loaded to a photonic chip. Each mode is coupled to a single-mode on-chip waveguide. The on-chip photonic circuit (iPronics Smartlight Processor) comprises 72~programmable MZIs arranged in a hexagonal mesh [Fig.~\ref{fig:GeneralConcept}(b)], with each MZI formed of phase shifters and symmetric mode couplers. The on-chip circuit is activated by external electrical control signals that drive the on-chip phase shifters to implement prescribed unitary and non-unitary operations on the $N$~modes, thus yielding an output field characterized by a new coherence matrix $\mathbf{G}'$. The modal weights (the diagonal elements of $\mathbf{G'}$) are measured by optical detectors, which may be either on-chip or off-chip. The electrical signals produced by the detectors are sent to an electronic processing unit (EPU), which also may be on-chip (a hybrid electronic-photonic chip) or off-chip. The EPU stores the measurements, and prepares electrical control signals that are fed back to the on-chip operation before measuring the modal weights again. In general, the electrical control signals that determine the structure of the photonic circuit at step $j+1$ may depend on the measured modal weights in the previous step $j$ (or all previous steps) or they may be predetermined, whereby the measurement sequence is fixed and each step is independent of the measured outcomes in previous steps.

\begin{figure*}[t!]
\centering
\includegraphics[width=16cm]{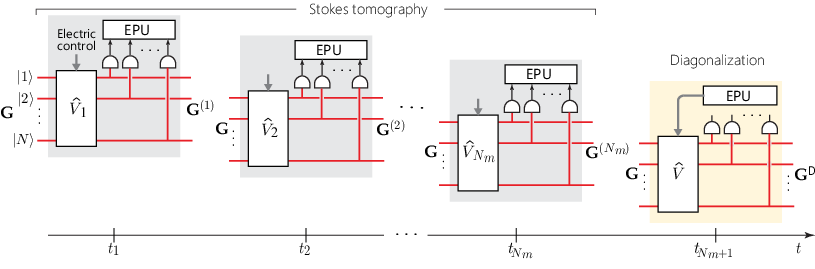}
\caption{On-chip diagonalization of a coherence matrix $\mathbf{G}$ spanned by $N$~modes via Stokes tomography. A fixed, predetermined sequence of $N_{\mathrm{m}}$ Stokes-tomography unitaries $\{\hat{V}_{1},\hat{V}_{2},\cdots,\hat{V}_{N_{\mathrm{m}}}\}$ are implemented at the time instances $\{t_{1},t_{2},\cdots,t_{N_{\mathrm{m}}}\}$. In any step $j$, where $j=1,2,\cdots,N_{\mathrm{m}}$, the unitary $\hat{V}_{j}$ is implemented, the resulting field is $\mathbf{G}^{(j)}=\hat{V}_{j}\mathbf{G}\hat{V}_{j}^{\dagger}$, and its modal weights are denoted $\{I_{1}^{(j)},I_{2}^{(j)},\cdots,I_{N}^{(j)}\}$, which are recorded by detectors and stored in an electronic processing unit (EPU). After the final Stokes-tomography unitary $\hat{V}_{N_{\mathrm{m}}}$ is implemented at time $t_{N_{\mathrm{m}}}$, the previously recorded measurements (from steps~1 through~$N_{\mathrm{m}}$) are combined to compute a final unitary $\hat{V}$ that is implemented at $t_{N_{\mathrm{m}}+1}$ to diagonalize $\mathbf{G}$ and produce $\mathbf{G}^{\mathrm{D}}=\hat{V}\mathbf{G}\hat{V}^{\dagger}$. The process of diagonalization via Stokes tomography thus takes $N_{\mathrm{m}}+1$~steps to complete.}
\label{fig:DiagonalizationProcedure}
\end{figure*}

\subsection{Structured coherence and the coherence matrix}

Structured coherence refers to a partially coherent optical field spanned by a finite number of modes \cite{Abouraddy26OPN,Abouraddy26AOP}. The modes are fixed, deterministic field distributions. When the modal coefficients are deterministic, the field is  coherent and represented by an $N\times1$ vector. When the modal coefficients are random variables, the field is partially coherent and described by an $N\times N$ Hermitian, positive semi-definite coherence matrix $\mathbf{G}$:
\begin{equation}
\mathbf{G}=\left(\begin{array}{cccc}
G_{11}&G_{12}&\cdots& G_{1,N}\\
G_{21}&G_{22}&\cdots&G_{2,N}\\
\vdots&\vdots&\ddots&\vdots\\
G_{N,1}&G_{N,2}&\cdots&G_{N,N}
\end{array}\right),
\end{equation}
where $G_{jk}=\langle E_{j}E_{k}^{*}\rangle=G_{kj}^{*}$, $j,k=1,\cdots,N$, $E_{j}$ is the field amplitude associated with the $j^{\mathrm{th}}$~mode labeled $|j\rangle$ in the Dirac notation \cite{Abouraddy26AOP}, and $\langle\cdot\rangle$ is an ensemble average. We normalize $\mathbf{G}$ to unity trace, $\mathrm{Tr}\{\mathbf{G}\}=\sum_{j=1}^{N}G_{jj}=1$. Because $\mathbf{G}$ is Hermitian, it is uniquely identified by $N^{2}$ real parameters reduced to $N^{2}-1$ once the unity-trace normalization is accounted for. The diagonal elements are `modal weights': the fractions of total power associated with the modes \cite{Abouraddy26AOP}. Modal detectors therefore register the diagonal elements of $\mathbf{G}$, whereas the off-diagonal elements of $\mathbf{G}$ cannot be measured directly. The coherence matrix can always be diagonalized via an appropriate $N\times N$ unitary: $\mathbf{G}=\hat{U}\mathbf{G}^{\mathrm{D}}\hat{U}^{\dagger}$, and $\mathbf{G}^{\mathrm{D}}=\hat{V}\mathbf{G}\hat{V}^{\dagger}$, where $\hat{U}$ and $\hat{V}$ are $N\times N$ unitaries, $\hat{V}=\hat{U}^{\dagger}$, and:
\begin{equation}
\mathbf{G}^{\mathrm{D}}=\left(\begin{array}{cccc}
\lambda_{1}&0&\cdots&0\\
0&\lambda_{2}&\cdots&0\\
\vdots&\vdots&\ddots&\vdots\\
0&0&\cdots&\lambda_{N}
\end{array}\right)=\mathrm{diag}\{\lambda_{1},\lambda_{2},\cdots,\lambda_{N}\},    
\end{equation}
$\{\lambda_{j}\}$ are the eigenvalues of $\mathbf{G}$, $\lambda_{j}\geq0$, $\sum_{j=1}^{N}\lambda_{j}=1$, and $\mathrm{diag}\{\cdot\}$ refers to a diagonal matrix with the entries corresponding to the diagonal elements \cite{Gamo64PO}. 

\subsection{Stokes Tomography}

A useful parameterization of a coherence matrix $\mathbf{G}$ makes use of the modal SPs $\{s_{j}\}$, which are the expansion coefficients of $\mathbf{G}$ in a matrix basis $\{\hat{\Lambda}_{j}\}$ \cite{Kimura03PLA,Bertlmann08JPA}, 
\begin{equation}\label{eq:GinTermsOfSPs}
\mathbf{G}=\frac{1}{2}\sum_{j=0}^{N^{2}-1}s_{j}\hat{\Lambda}_{j},
\end{equation}
and $\hat{\Lambda}_{0}=\hat{\mathbb{I}}_{N}$ is the $N\times N$ identity matrix. The matrices $\{\hat{\Lambda}_{j}\}$ are: (1) Hermitian, so that the modal SPs $\{s_{j}\}$ are real; (2) zero-trace $\mathrm{Tr}\{\hat{\Lambda}_{j}\}=0$, $j\neq0$; and (3) $\mathrm{Tr}\{\hat{\Lambda}_{j}\hat{\Lambda}_{k}\}=2\delta_{jk}$, $j,k=1,\cdots,N^{2}-1$. These properties allow us to extract the modal SPs via the projection $s_{j}=\mathrm{Tr}\{\hat{\Lambda}_{j}\mathbf{G}\}$. 

Because the number of SPs is $N^{2}-1$, it is commonly thought that $\mathcal{O}(N^{2})$ measurements are required for Stokes tomography \cite{Mor26arxiv}, which is not quite correct. Indeed, we show below that only $\mathcal{O}(N)$ measurements are needed as long as the $N$ modal weights are all recorded in each step. 

To reconstruct $\mathbf{G}$ via Stokes tomography, a predetermined, \textit{fixed} sequence of $N_{\mathrm{m}}$ steps is implemented. In each step, the following procedure is followed [Fig.~\ref{fig:DiagonalizationProcedure}]:
\begin{enumerate}
\item The on-chip photonic circuit is configured to realize one of $N_{\mathrm{m}}$ pre-determined unitaries $\{\hat{V}_{j}\}$, $j=1,2,\cdots,N_{\mathrm{m}}$, which transforms $\mathbf{G}$ at the input to $\mathbf{G}^{(j)}=\hat{V}_{j}\mathbf{G}\hat{V}_{j}^{\dagger}$ at the output. 
\item The modal weights associated with $\mathbf{G}^{(j)}$, denoted $\{I_{1}^{(j)},I_{2}^{(j)},\cdots,I_{N}^{(j)}\}$, are measured (corresponding to the diagonal elements of $\mathbf{G}^{(j)}$) and are stored in the EPU.
\end{enumerate}
After this sequence of $N_{\mathrm{m}}$ steps is concluded, occupying the time instances $\{t_{1},t_{2},\cdots,t_{N_{\mathrm{m}}}\}$, the stored measurements are used to calculate a final unitary $\hat{V}$ that is implemented by the photonic circuit at the instance $t_{N_{\mathrm{m}}+1}$. The unitary $\hat{V}$ is that which diagonalizes the input coherence matrix $\mathbf{G}$ and is obtained computationally from the reconstructed $\mathbf{G}$. By implementing $\hat{V}$, the output field is transformed into the diagonal coherence matrix $\hat{V}\mathbf{G}\hat{V}^{\dagger}=\mathbf{G}^{\mathrm{D}}$. Therefore, the number of steps needed to diagonalize $\mathbf{G}$ is $N_{\mathrm{m}}+1$ [Fig.~\ref{fig:DiagonalizationProcedure}].

We assume that $\mathbf{G}$ is constant during the $N_{\mathrm{m}}$~steps required for diagonalization. Furthermore, the on-chip MZI mesh allows for dynamically splitting the optical power in the initial $N$~modes, so that a copy of $\mathbf{G}$ is sent to the chip output or is made available on chip to carry out alternative processing of the field \textit{during} the Stokes-tomography procedure. After the final step, both $\mathbf{G}$ and its diagonalized counterpart $\mathbf{G}^{\mathrm{D}}$ are thus available. Alternatively, all the power can be dedicated to producing $\mathbf{G}^{\mathrm{D}}$ if there is no further need for the original field $\mathbf{G}$. The structure of the unitaries $\hat{V}_{j}$ and the number of required measurements $N_{\mathrm{m}}$ in relation to $N$~will become clear below.

\section{Diagonalization for two-mode light}

We describe in this Section the on-chip diagonalization of a $2\times2$ coherence matrix $\mathbf{G}$ for two-mode ($N=2$) light. The process starts by preparing off-chip generic two-mode incoherent light described by the coherence matrix $\mathbf{G}_{\mathrm{o}}=\tfrac{1}{2}\hat{\mathbb{I}}_{2}$, corresponding to uncorrelated, equal-amplitude modes ($\hat{\mathbb{I}}_{2}$ is the $2\times2$ identity matrix). This field is coupled to the chip whereupon a prescribed coherence matrix $\mathbf{G}$ is prepared in a two-step procedure: tuning the two eigenvalues $\lambda_{1}$ and $\lambda_{2}$ followed by a general $2\times2$ unitary $\hat{U}$. Any $2\times2$ coherence matrix $\mathbf{G}$ can be synthesized in this fashion \cite{Abouraddy26AOP}. The on-chip coherence matrix $\mathbf{G}$ is then diagonalized via Stokes tomography.

\subsection{Preparation of two-mode incoherent light}

\begin{figure}[t!]
\centering
\includegraphics[width=8.5cm]{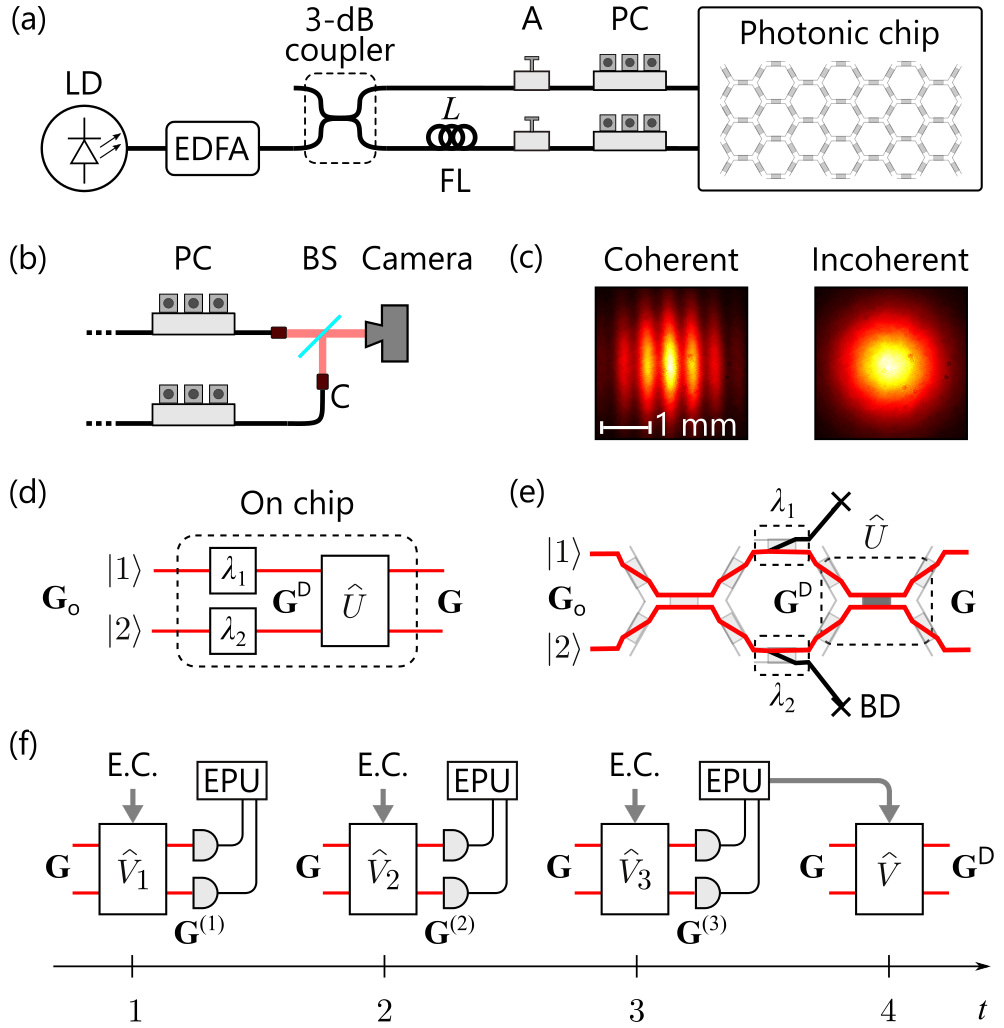} 
\caption{(a) Optical setup to prepare two mutually uncorrelated optical modes $|1\rangle$ and $|2\rangle$ and launch them into the photonic chip; LD: laser diode, EDFA: erbium-doped fiber amplifier, FL: fiber loop, A: attenuator, PC: polarization controller. (b) Setup to test the mutual coherence of modes $|1\rangle$ and $|2\rangle$. The two spatial modes are out-coupled via fiber collimators (C), combined by a symmetric beam splitter (BS), and directed together to a camera. (c) An interferogram is observed in absence of the fiber loop (coherent, high-visibility interference fringes, left panel) and \textit{not} in its presence (incoherent, no interference fringes, right panel). (d) Block diagram of the tasks implemented on chip to synthesize a prescribed coherence matrix $\mathbf{G}$. The eigenvalues $\lambda_{1}$ and $\lambda_{2}$ of the input coherence matrix $\mathbf{G}_{\mathrm{o}}=\tfrac{1}{2}\hat{\mathbb{I}}_{2}$ are tuned to yield $\mathbf{G}^{\mathrm{D}}$, and a $2\times2$ unitary $\hat{U}$ produces the coherence matrix $\mathbf{G}$. (e) On-chip circuit layout corresponding to the block diagram in (d) on the hexagonal MZI mesh to convert $\mathbf{G}_{\mathrm{o}}$ into $\mathbf{G}$; BD: beam dump. (f) Diagonalizing $\mathbf{G}$ via Stokes tomography. Three predetermined unitaries are implemented sequentially ($\hat{V}_{1}$, $\hat{V}_{2}$, and $\hat{V}_{3}$; Eq.~\ref{eq:StokesUnitaries}), after which the modal weights are recorded and stored in the electronic processing unit (EPU). After these three steps, $\mathbf{G}$ is reconstructed and the diagonalizing unitary $\hat{V}$ is calculated (Eq.~\ref{eq:2x2V}). In the fourth step of the procedure, $\hat{V}$ is implemented on chip to diagonalize $\mathbf{G}$.}
\label{fig:TwoModeSetup}
\end{figure}

The setup for preparing two-mode incoherent light is depicted schematically in Fig.~\ref{fig:TwoModeSetup}(a). Starting with an SMF-coupled laser diode at a wavelength of $\sim1550$~nm, we split the field into two SMFs (each representing one mode) via a 3-dB coupler. At this point, the modes are mutually correlated because they are derived from the same source. To render them mutually uncorrelated, we insert in one of the two paths a fiber loop of length $L\approx500$~m (exceeding the coherence length of the source $L_{\mathrm{o}}$). We place in each modal path an attenuator and a polarization controller to balance the power levels and ensure efficient coupling to the chip. The coherence matrix is $\mathbf{G}_{\mathrm{o}}=\tfrac{1}{2}\left(\!\!\begin{array}{cc}1&1\\1&1\end{array}\!\!\right)$ in absence of the fiber loop, corresponding to correlated and equal-amplitude modes, and $\mathbf{G}_{\mathrm{o}}=\tfrac{1}{2}\hat{\mathbb{I}}_{2}$
in presence of the fiber loop, corresponding to an incoherent field comprising two equal-amplitude, uncorrelated modes.

We confirm both field configurations (coherent and incoherent) by out-coupling the fields from the two SMFs via fiber collimators, before overlapping the two collimated fields at a balanced beam splitter, and recording the spatial intensity profile at a camera [Fig.~\ref{fig:TwoModeSetup}(b)]. In absence of the fiber loop, the field is \textit{coherent} and high-visibility interference fringes are observed, whereas no fringes are observed in presence of the fiber loop, indicating that the field has been rendered \textit{incoherent} [Fig.~\ref{fig:TwoModeSetup}(c)]. The field coupled to the chip is thus described by the $2\times2$ coherence matrix $\mathbf{G}_{\mathrm{o}}=\tfrac{1}{2}\hat{\mathbb{I}}_{2}$. 

\subsection{Preparing a prescribed coherence matrix on chip}

\begin{figure*}[t!]
\centering
\includegraphics[width=18cm]{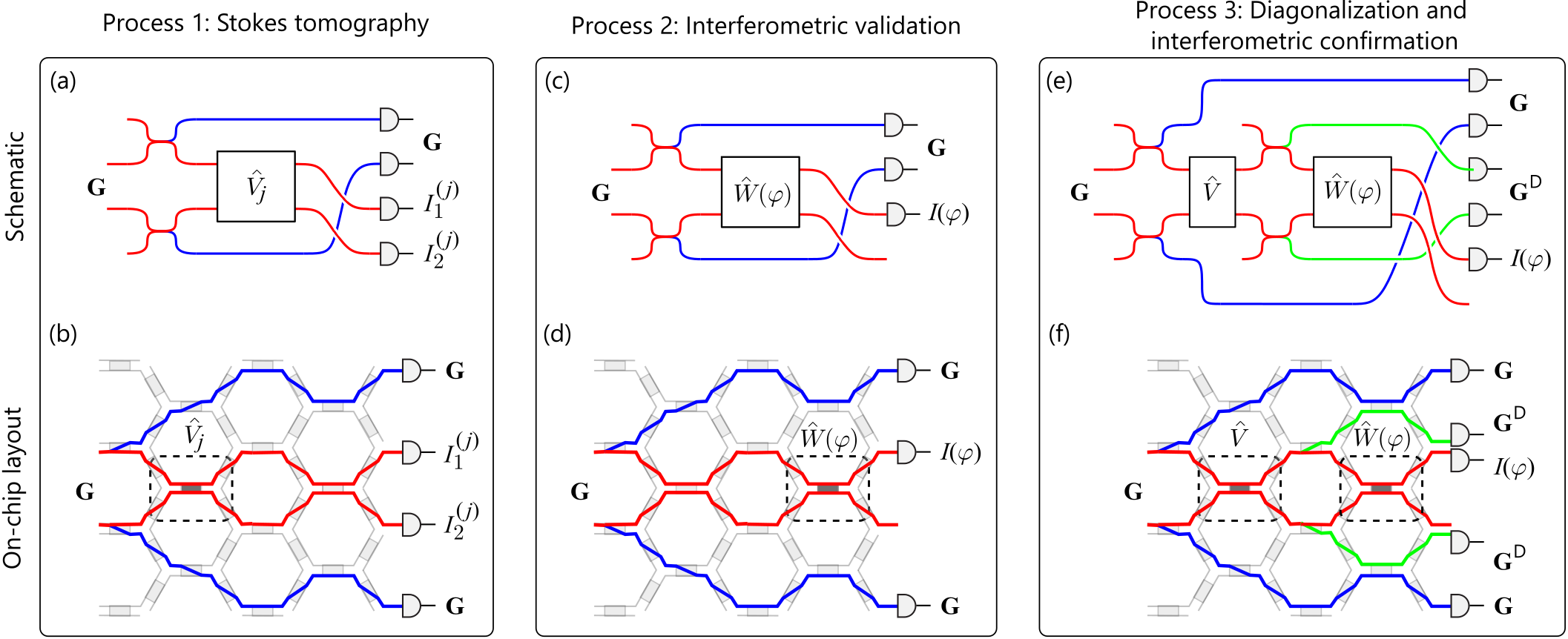} 
\caption{Photonic circuits for the on-chip processing tasks implemented on two-mode light to verify diagonalization. The top row shows schematics and the bottom row are the corresponding photonic-circuit layouts on the on-chip hexagonal mesh of MZIs. A portion of the modes $|1\rangle$ and $|2\rangle$ underpinning $\mathbf{G}$ are directed via couplers to the chip output to be made available throughout for independent processing. (a,b) Stokes-tomography unitaries $\hat{V}_{j}$, $j=1,2,3$, are implemented, and for we record for each the modal weights $I_{1}^{(j)}$ and $I_{2}^{(j)}$. (c,d) Interferometric validation that the modes underpinning $\mathbf{G}$ are partially correlated. We superpose the modes with a swept relative phase via the unitary $\hat{W}(\varphi)$, and record an output modal weight $I(\varphi)$. (e,f) Diagonalization. After reconstructing $\mathbf{G}$, we compute the unitary $\hat{V}$ and implement it on chip to produce the diagonalized coherence matrix $\mathbf{G}^{\mathrm{D}}$. We divide $\mathbf{G}^{\mathrm{D}}$ into two copies, one is directed to the chip output (where we measure the modal weights to confirm that they correspond to the eigenvalues of $\mathbf{G}$), and the other is directed to an interferometric unitary $\hat{W}(\varphi)$ to confirm that the modes underpinning $\mathbf{G}^{\mathrm{D}}$ are uncorrelated. The input coherence matrix $\mathbf{G}$ is still independently available at the output.}
\label{fig:TwoModeCircuits}
\end{figure*}

Once the two-mode generic field $\mathbf{G}_{\mathrm{o}}$ is coupled to the photonic chip, we tune the eigenvalues $\lambda_{1}$ and $\lambda_{2}$ of $\mathbf{G}_{\mathrm{o}}$ [Fig.~\ref{fig:TwoModeSetup}(d)] to adjust the field entropy $S=-\lambda_{1}\log_{2}\lambda_{1}-\lambda_{2}\log_{2}\lambda_{2}$ and its degree of coherence $D=\lambda_{1}-\lambda_{2}$ ($\lambda_{1}\geq\lambda_{2}$, $\lambda_{1}+\lambda_{2}=1$) \cite{Wolf07Book,Halder21OL,Abouraddy26AOP,Hashemi26arxivTwoMode,Abouraddy19Optica}. Although this is a \textit{non}-unitary process, it can nevertheless be accomplished on-chip in a unitary MZI by providing a zero field to one input MZI port and then ignoring an output port (whose field is guided to the edge of the chip to avoid crosstalk) [Fig.~\ref{fig:TwoModeSetup}(e)]. The MZI in this case effectively acts as a tunable on-chip attenuator, yielding a diagonal coherence matrix $\mathbf{G}^{\mathrm{D}}=\left(\begin{array}{cc}\lambda_{1}&0\\0&\lambda_{2}\end{array}\right)$. We produce three such fields:
\begin{equation}
\mathbf{G}_{1}^{\mathrm{D}}=\left(\!\begin{array}{cc}1&0\\0&0\end{array}\!\right)\!,
\mathbf{G}_{2}^{\mathrm{D}}=\left(\!\begin{array}{cc}0.8&0\\0&0.2\end{array}\!\right)\!,
\mathbf{G}_{3}^{\mathrm{D}}=\frac{1}{3}\left(\!\begin{array}{cc}2&0\\0&1\end{array}\!\right)\!;
\end{equation}
here $\mathbf{G}_{1}^{\mathrm{D}}$ is coherent ($D=1$ and $S=0$) while $\mathbf{G}_{2}^{\mathrm{D}}$ and $\mathbf{G}_{3}^{\mathrm{D}}$ are partially coherent: for $\mathbf{G}_{2}$, $D=0.6$ and $S\approx0.722$~bits; and for $\mathbf{G}_{3}$, $D\approx0.33$ and $S\approx0.91$~bits. We then realize a prescribed coherence matrix structure via a unitary. We implemented 4~unitaries $\{\hat{U}_{k}\}$, $k=1,\cdots,4$:
\begin{eqnarray}\label{eq:2x2Unitaries}
\hat{U}_{1}&=&-\frac{1}{\sqrt{2}}\left(\!\!\begin{array}{cc}1&-1\\1&1\end{array}\!\!\right)\!,
\hat{U}_{2}=-\frac{1}{\sqrt{2}}\left(\!\!\begin{array}{cc}1&1\\-1&1\end{array}\!\!\right)\!,\nonumber\\
\hat{U}_{3}&=&\frac{e^{i\tfrac{\pi}{4}}}{\sqrt{2}}\left(\!\!\begin{array}{cc}1&1\\i&-i\end{array}\!\!\right)\!,
\hat{U}_{4}=-\frac{e^{i\tfrac{\pi}{4}}}{\sqrt{2}}\left(\!\!\begin{array}{cc}1&1\\-i&i\end{array}\!\!\right)\!,
\end{eqnarray}
thus yielding $3\times4=12$ coherence matrices $\mathbf{G}_{j}^{(k)}=\hat{U}_{k}\mathbf{G}_{j}^{\mathrm{D}}\hat{U}_{k}^{\dagger}$. 

\subsection{Two-mode Stokes tomography}

Once $\mathbf{G}_{j}^{(k)}$ is prepared, we utilize Stokes tomography to reconstruct the coherence matrix and  diagonalize it [Fig.~\ref{fig:TwoModeSetup}(f)]. For $N=2$, the set $\{\hat{\Lambda}_{j}\}$ simply comprises the Pauli matrices $\{\hat{\sigma}_{j}\}$,
\begin{equation}\label{eq:Pauli}
\hat{\sigma}_{1}=\left(\begin{array}{cc}0&1\\1&0\end{array}\right),\;\hat{\sigma}_{2}=\left(\begin{array}{cc}0&-i\\i&0\end{array}\right),\;\hat{\sigma}_{3}=\left(\begin{array}{cc}1&0\\0&-1\end{array}\right),
\end{equation}
with $\hat{\sigma}_{0}=\hat{\mathbb{I}}_{2}$; $\hat{\sigma}_{1}$ and $\hat{\sigma}_{2}$ are off-diagonal matrices, whereas $\hat{\sigma}_{3}$ is a diagonal matrix. A generic $2\times2$ coherence matrix therefore takes the form:
\begin{equation}\label{eq:2x2Stokes}
\mathbf{G}=\left(\!\!\begin{array}{cc}G_{11}&G_{12}\\G_{21}&G_{22}\end{array}\!\!\right)=\sum_{j=0}^{3}s_{j}\hat{\sigma}_{j}=\frac{1}{2}\left(\!\!\begin{array}{cc}1+s_{3}&s_{1}-is_{2}\\s_{1}+is_{2}&1-s_{3}\end{array}\!\!\right)\!.
\end{equation}
The procedure to reconstruct $\mathbf{G}$ occurs in three `steps', where a `step' encompasses changing the on-chip photonic circuit via an external electrical signal and recording the modal weights $G_{11}$ and $G_{22}$. We implement Stokes-tomography unitaries $\hat{V}_{j}$:
\begin{equation}\label{eq:StokesUnitaries}
\hat{V}_{1}=\frac{1}{\sqrt{2}}\left(\!\begin{array}{cc}
1&1\\-1&1\end{array}\!\right)\!,\;
\hat{V}_{2}=\frac{1}{\sqrt{2}}\left(\!\begin{array}{cc}
1&-i\\-i&1\end{array}\!\right)\!,\;
\hat{V}_{3}=\hat{\mathbb{I}}_{2}.
\end{equation}
The unitaries $\hat{V}_{1}$ and $\hat{V}_{2}$ are associated with the off-diagonal element of $\mathbf{G}$. In general, after implementing $\hat{V}_{j}$, we record the modal weights $I_{1}^{(j)}=\tfrac{1}{2}(1+s_{j})$ and $I_{2}^{(j)}=\tfrac{1}{2}(1-s_{j})$, and obtain their difference $\Delta_{j}=I_{1}^{(j)}-I_{2}^{(j)}\propto s_{j}$. In the first step, we implement $\hat{V}_{1}$ and record $I_{1}^{(1)}$ and $I_{2}^{(1)}$, from which we obtain $s_{1}$, and in the second step implement $\hat{V}_{2}$ and record $I_{2}^{(1)}$ and $I_{2}^{(2)}$, from which we obtain $s_{2}$. In the third step, to obtain the diagonal elements $G_{11}$ and $G_{22}$, we measure the modal weights directly corresponding to the field traversing $\hat{V}_{3}=\hat{\mathbb{I}}_{2}$ [Fig.~\ref{fig:TwoModeSetup}(f)]. These three measurements provide $s_{1}$, $s_{2}$, and $s_{3}$, from which we can reconstruct $\mathbf{G}$ via Eq.~\ref{eq:2x2Stokes}, and then calculate the $2\times2$ unitary $\hat{V}$ that diagonalizes $\mathbf{G}$. Alternatively, we can directly calculate $\hat{V}$ without reconstructing $\mathbf{G}$ using the SPs via:
\begin{equation}\label{eq:2x2V}
\hat{V}=\frac{1}{\sqrt{2D(D-s_{3})}}\left(\begin{array}{cc}s_{1}+is_{2}&D-s_{3}\\s_{3}-D&s_{1}-is_{2}\end{array}\right),
\end{equation}
where $D=\sqrt{s_{1}^{2}+s_{2}^{2}+s_{3}^{2}}$. In a fourth step [Fig.~\ref{fig:TwoModeSetup}(f)], we implement the unitary $\hat{V}$ in place of the Stokes-tomography unitaries $\hat{V}_{j}$, after which the output field is $\mathbf{G}^{\mathrm{D}}=\left(\begin{array}{cc}\lambda_{1}&0\\0&\lambda_{2}\end{array}\right)$, and Stokes-tomography-based diagonalization is completed.

\begin{table*}[t!]
\caption{Parameters characterizing the~12 synthesized coherence matrices $\mathbf{G}_{j}^{(k)}$, $j=1,2,3$, and $k=1,\cdots,4$. We list values of the reconstruction fidelity $\mathcal{F}$, the theoretical values of entropy $S_{\mathrm{th}}$, the entropy calculated from the reconstructed coherence matrix $S_{\mathrm{meas}}$, and the entropy calculated from the diagonalized modal weights $S_{\mathrm{D}}$. We also list the theoretical value of the degree of coherence $D$, the observed interference-fringe visibility $V_{\mathrm{before}}$ produced by the original two modes, and that observed after diagonalization $V_{\mathrm{D}}$.}\label{Table:2x2}

\setlength\extrarowheight{1pt}

\begin{tabular}
{|c||c|c|c|c||c|c|c|c||c|c|c|c|}
\hline
& $G_{1}^{(1)}$ & $G_{1}^{(2)}$ & $G_{1}^{(3)}$ & $G_{1}^{(4)}$ & $G_{2}^{(1)}$ & $G_{2}^{(2)}$ & $G_{2}^{(3)}$ & $G_{2}^{(4)}$ & $G_{3}^{(1)}$ & $G_{3}^{(2)}$ & $G_{3}^{(3)}$ & $G_{3}^{(4)}$
 
\\\hline\hline

$\mathcal{F}$ &0.99&0.95&0.98&0.99&0.99&0.99&0.99&0.99&0.99&0.98&0.99&0.98
\\\hline

$S_{\mathrm{th}}$ &0&0&0&0&0.722&0.722&0.722&0.722&0.91&0.91&0.91&0.91
\\\hline

$S_{\mathrm{meas}}$ &0.11&0.29&0.14&0.10&0.639&0.797&0.7242&0.614&0.869&0.956&0.912&0.855
\\\hline

$S_{\mathrm{D}}$ & 0.136&0.106&0.124&0.119&0.681&0.711&0.634&0.711&0.881&0.899&0.856&0.913
\\\hline

$D$ & 1&1&1&1&0.6&0.6&0.6&0.6&0.333&0.333&0.333&0.333
\\\hline

$V_{\mathrm{before}}$ &0.975&0.984&0.97&0.976&0.644&0.622&0.654&0.625&0.391&0.37&0.401&0.36
\\\hline

$V_{\mathrm{D}}$ &0.017&0.069&0.064&0.042&0.033&0.08&0.033&0.026&0.04&0.08&0.038&0.022

\\
\hline

\end{tabular}
\end{table*}

\subsection{On-chip processing and verification tasks}

We have outlined above the conceptual scheme for diagonalizing a bimodal field via Stokes tomography [Fig.~\ref{fig:TwoModeSetup}(f)].To verify the diagonalization procedure, we implement three processes [Fig.~\ref{fig:TwoModeCircuits}]. We first make use of two couplers to direct a portion of the modes $|1\rangle$ and $|2\rangle$ supporting $\mathbf{G}$ to the chip output away from any subsequent system. Therefore, the input field $\mathbf{G}$ is made available independently of the Stokes tomography procedure. We next describe the sequence of measurement processes implemented on chip to validate this procedure.

\textit{Process~1: Stokes tomography}. In the first process, we implement the sequence of Stokes unitaries $\hat{V}_{j}$ [Fig.~\ref{fig:TwoModeCircuits}(a,b)] to reconstruct $\mathbf{G}$ and obtain the unitary $\hat{V}$ that diagonalizes $\mathbf{G}$.

\textit{Process~2: Interferometric validation of modal correlations}. In the second process, we direct $\mathbf{G}$ to a modal interferometer described by the unitary $\hat{W}(\varphi)$:
\begin{equation}\label{eq:Interferometer}
\hat{W}(\varphi)=\frac{e^{-i\varphi/2}}{\sqrt{2}}\left(\begin{array}{cc}e^{i\varphi}&1\\e^{i\varphi}&-1\\\end{array}\right), 
\end{equation}
where $\varphi$ is a swept phase and we record the modal weight $I(\varphi)$. If the two modes are mutually correlated ($\mathbf{G}$ is \textit{not} diagonal), then $I(\varphi)$ will display interference fringes; and if they are mutually uncorrelated ($\mathbf{G}$ is diagonal), then $I(\varphi)$ remains constant with $\varphi$. The interference visibility corresponds to the degree of coherence $D_{j}$ as long as the modal weights are equal. The unitaries $\hat{U}_{k}$ in Eq.~\ref{eq:2x2Unitaries} are all symmetrizing unitaries; i.e., the modal weights of the coherence matrix produced are equal, independently of the initial coherence matrix. We thus expect the observed visibility to be independent of $\hat{U}_{k}$. We sketch the block diagram for this procedure in Fig.~\ref{fig:TwoModeCircuits}(c) and the corresponding on-chip layout in Fig.~\ref{fig:TwoModeCircuits}(d).

\textit{Process~3: Diagonalization and verification of modal correlations}. In the third process, we implement $\hat{V}$ on $\mathbf{G}$ to produce $\mathbf{G}^{\mathrm{D}}$ [Fig.~\ref{fig:TwoModeCircuits}(e,f)]. To confirm that the modes underpinning $\mathbf{G}^{\mathrm{D}}$ are indeed uncorrelated, we split the modes of $\mathbf{G}^{\mathrm{D}}$ into two copies, one copy is routed to the output while the other copy traverses again the modal interferometer $\hat{W}(\varphi)$ in Eq.~\ref{eq:Interferometer}, after which we record $I(\varphi)$ [Fig.~\ref{fig:TwoModeCircuits}(e,f)]. We confirm that $\mathbf{G}$ has been diagonalized in two distinct ways: (1) measuring the modal weights and confirming that they correspond to the eigenvalues of $\mathbf{G}$; and (2) verifying interferometrically that the modes are uncorrelated through lack of fringe visibility while sweeping $\varphi$.

\subsection{Measurement results for two-mode diagonalization}

\begin{figure*}[p]
\centering
\includegraphics[width=15.5cm]{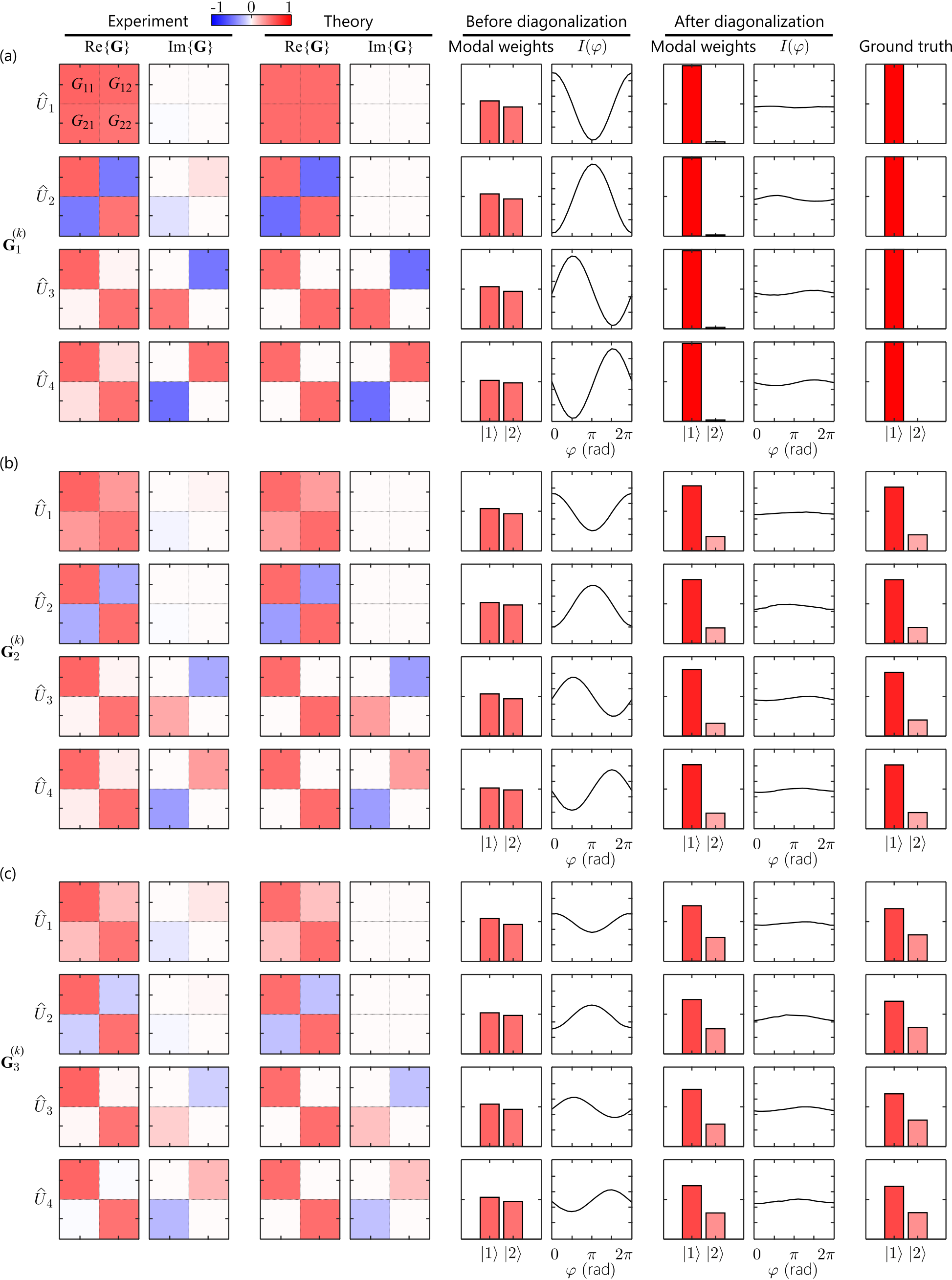} 
\caption{Stokes-tomography-based diagonalization for two-mode light. First and second columns are the real and imaginary parts of the reconstructed coherence matrices, whereas the third and and fourth columns correspond to the theoretically expected values. The fifth column shows the modal weights of the reconstructed coherence matrices. The sixth column depicts the measured interferogram $I(\varphi)$ resulting from superposing the two input modes with a swept relative phase $\varphi$. The seventh and eighth columns depict the modal weights and the interferogram $I(\varphi)$ for the field after diagonalization. The final column is the ground truth for the eigenvalues of the coherence matrices. The rows in each panel correspond to fields having the same entropy but differ in the unitary implemented in its synthesis: (a) $\mathbf{G}_{1}^{(k)}$ with $D=1$ and $S=0$; (b) $\mathbf{G}_{2}^{(k)}$ with $D=0.6$ and $S\approx0.722$~bits; and (c) $\mathbf{G}_{3}^{(k)}$ with $D\approx0.33$ and $S\approx0.91$~bits.}
\label{fig:TwoModeData}
\end{figure*}

We plot in Fig.~\ref{fig:TwoModeData} the measurement results arranged according to the three processes outlined above, and list the salient parameters extracted from the measurements in Table~\ref{Table:2x2}.

\textit{Process~1}. First, we plot the reconstructed real and imaginary parts of the elements of the 12~reconstructed $2\times2$ matrices $\mathbf{G}_{j}^{(k)}$ and compare them to theoretical predictions, which occupy the first 4 columns of Fig.~\ref{fig:TwoModeData}. The reconstructed matrices are obtained after the $N_{\mathrm{m}}=3$ steps that acquire the SPs. To benchmark the quality of the Stokes tomographic reconstruction, we make use of the fidelity $0\leq\mathcal{F}\leq1$, a metric we borrow from quantum information processing \cite{Jozsa94JMO}, which is defined as $\mathcal{F}=\mathrm{Tr}\{\sqrt{\mathbf{G}_{\mathrm{m}}}\mathbf{G}_{\mathrm{th}}\sqrt{\mathbf{G}_{\mathrm{m}}}\})^{2}$, where $\mathbf{G}_{\mathrm{th}}$ and $\mathbf{G}_{\mathrm{m}}$ are the theoretically expected and the measured coherence matrices, respectively. The values of $\mathcal{F}$ for the~12 reconstructed coherence matrices are listed in Table~\ref{Table:2x2}, and we find that $\mathcal{F}\geq95\%$ in all cases. The values of the entropy obtained from the reconstructed coherence matrices $S_{\mathrm{meas}}$ are also given in Table~\ref{Table:2x2} and are compared to the theoretically expected values of entropy $S_{\mathrm{th}}$. In general, lower values of $\mathcal{F}$ are consistently observed only in the cases where entropy is close to $S=0$, whereupon minute discrepancies lead to large errors in estimating $S$. This step verifies that we have reconstructed the coherence matrices correctly.

\textit{Process~2}. Here we direct the modes $|1\rangle$ and $|2\rangle$ of the input field to the modal interferometer $\hat{W}(\varphi)$ and plot the output $I(\varphi)$ from one port in Fig.~\ref{fig:TwoModeData}. Observing an interferogram confirms that the two modes are partially coherent, and the visibility is proportional to the amplitude of the off-diagonal element of the coherence matrix. In all cases, $I_{1}\approx I_{2}$, as expected from the symmetrizing form of the unitaries $\hat{U}_{k}$ in Eq.~\ref{eq:2x2Unitaries}. The measured interferograms confirm these expectations. For $\mathbf{G}_{1}^{(k)}$, $k=1,\cdots,4$, which correspond to coherent fields, the visibility is close to unity. This indicates that the two modes are fully correlated. As we reduce the degree of coherence, the interference visibility of $I(\varphi)$ drops accordingly. We list the measured values of visibility $V_{\mathrm{before}}$ in Table~\ref{Table:2x2} and observe -- as expected -- that they are independent of $k$ and match the degree of coherence $D_{j}$.

\textit{Process~3}. After reconstructing $\mathbf{G}_{j}^{(k)}$ in Process~!, we calculate the unitary $\hat{V}$ that diagonalizes it (Eq.~\ref{eq:2x2V}), and implement it on-chip to yield the diagonalized coherence matrix. We plot in Fig.~\ref{fig:TwoModeData} the modal weights after diagonalization, which are in excellent agreement with $I_{1}=G_{11}=\lambda_{1}$ and $I_{2}=G_{22}=\lambda_{2}$ for the 12~reconstructed coherence matrices. We calculate the entropies from the diagonalized coherence matrix $S_{\mathrm{D}}$ and list them in Table~\ref{Table:2x2}. After diagonalizing the coherence matrix, we direct the modes $|1\rangle$ and $|2\rangle$ again to the modal interferometer $\hat{W}(\varphi)$ and plot the output $I(\varphi)$ from one port in Fig.~\ref{fig:TwoModeData}. After diagonalization, the field is left in the coherent-mode representation, so that the modes $|1\rangle$ and $|2\rangle$ are uncorrelated, which is manifest here in the lack of interferometric fringes independently of the field entropy, thereby confirming that the two modes in all cases are now uncorrelated. This, along with the measured modal weights, verifies that the coherence matrix has been correctly diagonalized.

\section{Diagonalization for four-mode light}\label{sec:FourModes}

\subsection{Preparation of the four-mode field}

\begin{figure}[t!]
\centering
\includegraphics[width=8.5cm]{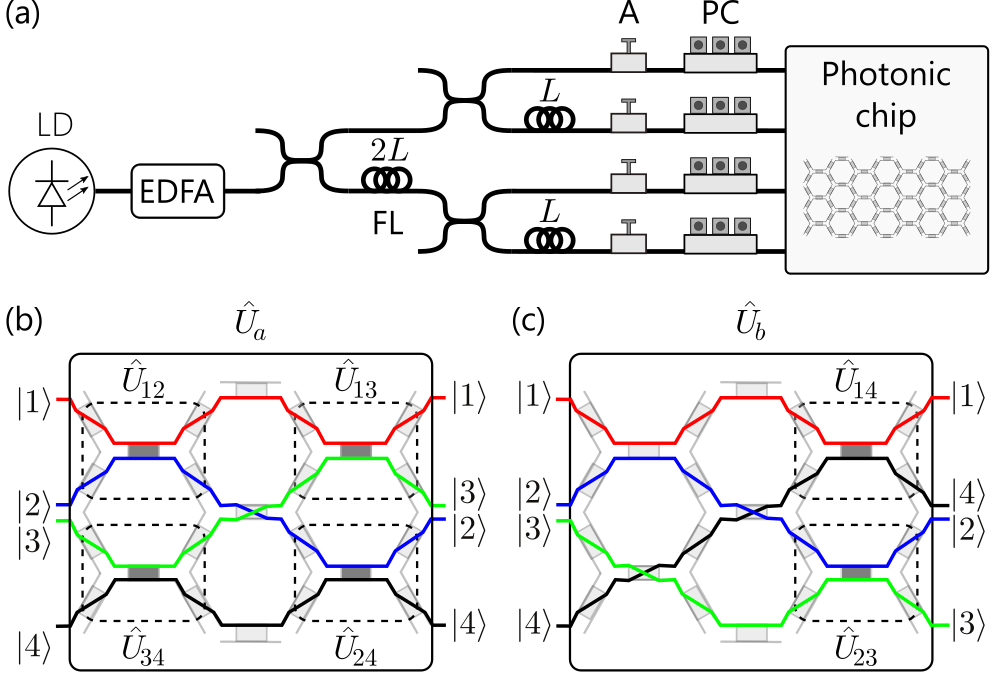} 
\caption{(a) Optical setup to prepare and launch the input field $\mathbf{G}_{\mathrm{o}}=\tfrac{1}{4}\hat{\mathbb{I}}_{4}$ comprising four uncorrelated, equal-power modes, and then tune the eigenvalues $\{\lambda_{j}\}$ to obtain the coherence matrices $\mathbf{G}^{\mathrm{D}}$, $j=1,\cdots,4$; LD: laser diode, FL: fiber loop, A: attenuator, and PC: polarization controller. (b,c) Photonic circuits to construct the $4\times4$ unitaries (b) $\hat{U}_{a}$ and (c) $\hat{U}_{b}$ out of a succession of $2\times2$ unitaries. The modal paths are identified with different colors; the order of the modes after the unitaries differs from that before them. The dark gray rectangles identify the $2\times2$ unitaries.}
\label{fig:FourModeSetup}
\end{figure}

We now move on to a demonstration of the on-chip, Stokes-tomographic diagonalization procedure for four-mode light ($N=4$). The setup for preparing a four-mode incoherent field is depicted in Fig.~\ref{fig:FourModeSetup}(a). We start with the same SMF-coupled laser diode employed in Fig.~\ref{fig:TwoModeSetup}(a), except that we now split the field equally into 4 paths via 3-dB couplers and insert appropriate fiber lengths in each path, such that any pair of paths differ by at least a length $L\approx500$~m ($L>L_{\mathrm{o}}$). Because the field amplitudes are equal and are uncorrelated, the coherence matrix representing the field is $\mathbf{G}_{\mathrm{o}}=\tfrac{1}{4}\hat{\mathbb{I}}_{4}$, where $\hat{\mathbb{I}}_{4}$ is the $4\times4$ identity matrix. We make use of fiber attenuators to adjust the modal weights to $\{\lambda_{1},\lambda_{2},\lambda_{3},\lambda_{4}\}$ and thus produce $\mathbf{G}^{\mathrm{D}}$. This enables tuning both the \textit{entropy} of $\mathbf{G}^{\mathrm{D}}$ and its \textit{rank}, which we have shown to be a useful parameter for scattering-free communications, in addition to underpinning a host of novel concepts in coherence theory \cite{Harling24PRA,Harling24PRA2,Harling25APLP}. We thus produce the diagonal coherence matrix:
\begin{equation}
\mathbf{G}^{\mathrm{D}}=\left(\begin{array}{cccc}
\lambda_{1}&0&0&0\\0&\lambda_{2}&0&0\\0&0&\lambda_{3}&0\\0&0&0&\lambda_{4}
\end{array}\right)=\mathrm{diag}\{\lambda_{1},\lambda_{2},\lambda_{3},\lambda_{4}\},
\end{equation}
with $\sum_{j=1}^{4}\lambda_{j}=1$. We produce a representative field for each rank (rank-1, rank-2, rank-3, and rank-4), given by:
\begin{eqnarray}
\mathbf{G}_{1}^{\mathrm{D}}&=&\mathrm{diag}\{1,0,0,0\},\nonumber\\
\mathbf{G}_{2}^{\mathrm{D}}&=&\mathrm{diag}\{0.7,0.3,0,0\},\nonumber\\
\mathbf{G}_{3}^{\mathrm{D}}&=&\mathrm{diag}\{0.5,0.3,0.2,0\},\nonumber\\
\mathbf{G}_{4}^{\mathrm{D}}&=&\mathrm{diag}\{0.4,0.3,0.2,0.1\}.
\end{eqnarray}
In addition to changing the rank of $\mathbf{G}$, this step also tunes its entropy $S=-\sum_{j=1}^{4}\lambda_{j}\log_{2}\lambda_{j}$; here $S=0$ for $\mathbf{G}_{1}$, here $S=0.88$ for $\mathbf{G}_{2}$, here $S=1.485$ for $\mathbf{G}_{3}$, and here $S=1.846$ for $\mathbf{G}_{4}$.

The fields from the 4~SMFs (corresponding to the coherence matrices $\mathbf{G}_{j}^{\mathrm{D}}$) are coupled to the chip, where the structure of the coherence matrix is sculpted via an on-chip $4\times4$ unitary. We have implemented two such unitaries $\hat{U}_{a}$ and $\hat{U}_{b}$. The first unitary $\hat{U}_{a}$ is given by: 
\begin{equation}
\hat{U}_{a}=\frac{1}{2}\left(\begin{array}{cccc}1&1&-i&i\\1&-1&-i&-i\\i&i&-1&1\\-i&i&1&1\end{array}\right),
\end{equation}
which is constructed out of a sequence of 4~unitaries $\hat{U}_{12}$, $\hat{U}_{34}$, $\hat{U}_{13}$, and $\hat{U}_{24}$, where the indices identify the pair of modes the unitary operates on [Fig.~\ref{fig:FourModeSetup}(b)]. These unitaries are given explicitly by:
\begin{eqnarray}
\hat{U}_{12}&=&\frac{1}{\sqrt{2}}\left(\begin{array}{cc}1&1\\1&-1\end{array}\right),
\hat{U}_{34}=\frac{1}{\sqrt{2}}\left(\begin{array}{cc}1&-1\\-1&-1\end{array}\right),\nonumber\\
\hat{U}_{13}&=&\frac{1}{\sqrt{2}}\left(\begin{array}{cc}1&-i\\i&-1\end{array}\right),
\hat{U}_{24}=\frac{1}{\sqrt{2}}\left(\begin{array}{cc}1&i\\-i&-1\end{array}\right).
\end{eqnarray}
The second $4\times4$ unitary $\hat{U}_{b}$ implemented is given by:
\begin{equation}
\hat{U}_{b}=\frac{1}{2}\left(\begin{array}{cccc}1&0&0&\sqrt{3}\eta^{2}\\0&\sqrt{3}&\eta&0\\0&\eta&-\sqrt{3}&0\\\sqrt{3}\eta^{-2}&0&0&-1\end{array}\right),
\end{equation}
where $\eta=e^{-i\pi/6}$. This unitary is constructed out of 2~unitaries $\hat{U}_{23}$ and $\hat{U}_{14}$ [Fig.~\ref{fig:FourModeSetup}(c)], given by:
\begin{equation}
\hat{U}_{23}=\frac{1}{2}\left(\!\!\begin{array}{cc}
\sqrt{3}&\eta\\\eta^{*}&-\sqrt{3}
\end{array}\!\!\right),
\hat{U}_{14}=\frac{1}{2}\left(\!\!\begin{array}{cc}
1&\sqrt{3}\eta^{2}\\\sqrt{3}\eta^{-2}&-1
\end{array}\!\!\right).
\end{equation}

Combining the four coherence matrices $\mathbf{G}_{j}^{\mathrm{D}}$ ($j=1,\cdots,4$) and the unitary $\hat{U}_{a}$ and $\hat{U}_{b}$, we prepare in total 8~coherence matrices: $\mathbf{G}_{j}^{(k)}=\hat{U}_{k}\mathbf{G}_{j}^{\mathrm{D}}\hat{U}_{k}^{\dagger}$, where $j=1,2,3,4$ and $k=a,b$.

\subsection{Generalized Pauli matrices and Stokes parameters for four-mode light}

The $4\times4$ coherence matrix $\mathbf{G}$ can be written as a superposition of a matrix basis $\mathbf{G}=\tfrac{1}{2}\sum_{j=0}^{15}s_{j}\hat{\Lambda}_{j}$, whose 16~expansion coefficients $\{s_{j}\}$ are the modal SPs, which are associated with $4\times4$ generalized Stokes matrices $\{\hat{\Lambda}_{j}\}$, including $\hat{\Lambda}_{0}=\hat{\mathbb{I}}_{4}$. These matrices share the properties listed earlier for Pauli matrices: (1) they are Hermitian (so that the SPs are real); (2) $\mathrm{Tr}\{\hat{\Lambda}_{j}\}=0$ for $j\neq0$; and (3) $\mathrm{Tr}\{\hat{\Lambda}_{j}\hat{\Lambda}_{k}\}=\delta_{j,k}$, $j,k=1,\cdots,15$. We divide these matrices into two subsets: 12~off-diagonal and 4~diagonal matrices. We need not define the diagonal matrices because the diagonal elements of $\mathbf{G}$ are acquired by recording the modal weights. We utilize the following set of off-diagonal matrices:
\begin{equation}
\hat{\Lambda}_{1}=\left(\begin{array}{cccc}
0&1&0&0\\1&0&0&0\\0&0&0&0\\0&0&0&0
\end{array}\right),\;
\hat{\Lambda}_{2}=\left(\begin{array}{cccc}
0&-i&0&0\\i&0&0&0\\0&0&0&0\\0&0&0&0
\end{array}\right),\nonumber
\end{equation}
\begin{equation}
\hat{\Lambda}_{3}=\left(\begin{array}{cccc}
0&0&1&0\\0&0&0&0\\1&0&0&0\\0&0&0&0
\end{array}\right),\;
\hat{\Lambda}_{4}=\left(\begin{array}{cccc}
0&0&-i&0\\0&0&0&0\\i&0&0&0\\0&0&0&0
\end{array}\right),\nonumber
\end{equation}
\begin{equation}
\hat{\Lambda}_{5}=\left(\begin{array}{cccc}
0&0&0&1\\0&0&0&0\\0&0&0&0\\1&0&0&0
\end{array}\right),\;
\hat{\Lambda}_{6}=\left(\begin{array}{cccc}
0&0&0&-i\\0&0&0&0\\0&0&0&0\\i&0&0&0
\end{array}\right),\nonumber
\end{equation}
\begin{equation}
\hat{\Lambda}_{7}=\left(\begin{array}{cccc}
0&0&0&0\\0&0&1&0\\0&1&0&0\\0&0&0&0
\end{array}\right),\;
\hat{\Lambda}_{8}=\left(\begin{array}{cccc}
0&0&0&0\\0&0&-i&0\\0&i&0&0\\0&0&0&0
\end{array}\right),\nonumber
\end{equation}
\begin{equation}
\hat{\Lambda}_{9}=\left(\begin{array}{cccc}
0&0&0&0\\0&0&0&1\\0&0&0&0\\0&1&0&0
\end{array}\right),\;
\hat{\Lambda}_{10}=\left(\begin{array}{cccc}
0&0&0&0\\0&0&0&-i\\0&0&0&0\\0&i&0&0
\end{array}\right),\nonumber
\end{equation}
\begin{equation}\label{eq:4x4Pauli}
\hat{\Lambda}_{11}=\left(\begin{array}{cccc}
0&0&0&0\\0&0&0&0\\0&0&0&1\\0&0&1&0
\end{array}\right),\;
\hat{\Lambda}_{12}=\left(\begin{array}{cccc}
0&0&0&0\\0&0&0&0\\0&0&0&-i\\0&0&i&0
\end{array}\right),\nonumber
\end{equation}

\begin{figure*}[t!]
\centering
\includegraphics[width=16.2cm]{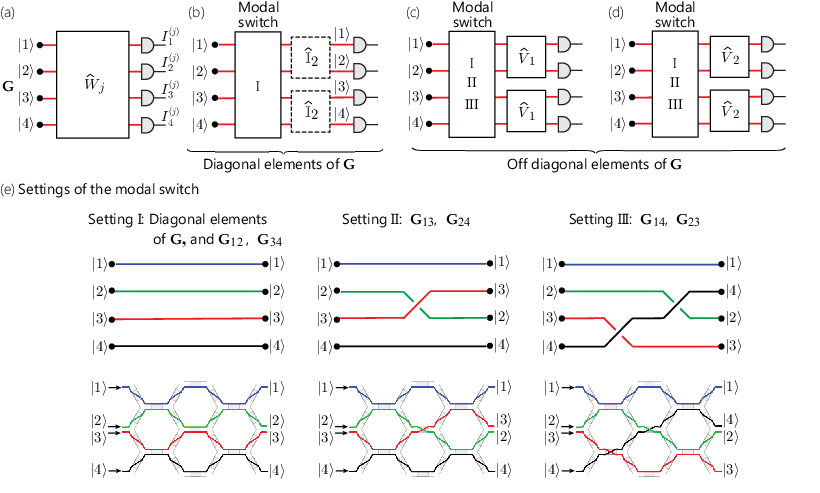}
\caption{Diagonalization procedure for four-mode light. (a) In each step, a $4\times4$ unitary $\hat{W}_{j}$ is implemented, and the modal weights are recorded. (b) Each unitary $\hat{W}_{j}$ is divided into a modal switch followed by two $2\times2$ unitaries implemented on two pairs of modes. To acquire the diagonal elements of $\mathbf{G}$, the modal switch is in setting~I and the modal weights are recorded. (c,d) To acquire the off-diagonal elements of $\mathbf{G}$, the modal switch takes on one of three settings (I, II, or III), and in each setting, two identical unitaries are implemented on pairs of modes: (c) $\hat{V}_{1}$ and $\hat{V}_{1}$, or (d) $\hat{V}_{2}$ and $\hat{V}_{2}$. (e) The three settings~I, II, and III of the modal switch. The modes are depicted by differently colored lines. The top row shows conceptual schematics and the bottom row shows on-chip photonic-circuit layouts. In setting~I, the modal order is unchanged; in setting~II, modes $|2\rangle$ and $|3\rangle$ are switched, corresponding to the reordering $\{|1\rangle,|2\rangle,|3\rangle,|4\rangle\}\rightarrow\{|1\rangle,|3\rangle,|2\rangle,|4\rangle\}$; and in setting~III, we have the reordering $\{|1\rangle,|2\rangle,|3\rangle,|4\rangle\}\rightarrow\{|1\rangle,|4\rangle,|2\rangle,|3\rangle\}$.}
\label{fig:4x4GeneralStokesRecon}
\end{figure*}

It is clear that each of these $4\times4$ matrices is simply either a $2\times2$ Pauli matrix $\hat{\sigma}_{1}$ or $\hat{\sigma}_{2}$ embedded in the four-mode space. For example, $\hat{\Lambda}_{1}$ has the Pauli matrix $\hat{\sigma}_{1}$ implemented on modes $|1\rangle$ and $|2\rangle$, and thus identifies the real part of the off-diagonal element $G_{12}$ of $\mathbf{G}$, whereas the matrix $\hat{\Lambda}_{2}$ has the Pauli matrix $\hat{\sigma}_{2}$ implemented on modes $|1\rangle$ and $|2\rangle$, and thus identifies the imaginary part of $G_{12}$. Similarly, $\hat{\Lambda}_{3}$ and $\hat{\Lambda}_{4}$ have the Pauli matrices $\hat{\sigma}_{1}$ and $\hat{\sigma}_{2}$, respectively, implemented on modes $|1\rangle$ and $|3\rangle$, and thus identify the real and imaginary parts of the off-diagonal element $G_{13}$ of $\mathbf{G}$, respectively; and so on.

\subsection{Reconstructing the coherence matrix for four-mode light}

One may of course proceed to implement~12 unitaries $\hat{W}_{j}$, $j=1,2,\cdots,12$ [Fig.~\ref{fig:4x4GeneralStokesRecon}(a)] on the four modes, each related to one of the off-diagonal generalized Pauli matrices $\{\hat{\Lambda}_{j}\}$ listed above. The unitaries $\hat{W}_{j}$ are all $4\times4$. However, the structure of the generalized Pauli matrices $\{\hat{\Lambda}_{j}\}$ suggests a simplification of the implementations of $\hat{W}_{j}$ that reduces the total number of steps required to reconstruct $\mathbf{G}$. 

We first measure the \textit{diagonal} elements of $\mathbf{G}$ by simply detecting the modal weights associated with the modes $\{|1\rangle,|2\rangle,|3\rangle,|4\rangle\}$ [Fig.~\ref{fig:4x4GeneralStokesRecon}(b)]. Next, to acquire the \textit{off-diagonal} elements of $\mathbf{G}$. We divide the unitary $\hat{W}_{j}$ into two stages. In the first stage, a modal switch is implemented that re-orders the modes. In the second stage, $2\times2$ unitaries are applied to pairs of modes; either (1) $\hat{V}_{1}$ (Eq.~\ref{eq:StokesUnitaries}) on two modes \textit{and} $\hat{V}_{1}$ on the remaining two modes, which allows acquiring the real parts of two off-diagonal elements of $\mathbf{G}$; \textit{or} (2) $\hat{V}_{2}$ (Eq.~\ref{eq:StokesUnitaries}) on the two modes \textit{and} $\hat{V}_{2}$ on the remaining two modes, which allows acquiring the imaginary parts of the same two off-diagonal elements of $\mathbf{G}$. We thus acquire two complex off-diagonal elements of $\mathbf{G}$ in two steps rather than in 4~steps [Fig.~\ref{fig:4x4GeneralStokesRecon}(c,d)].

\begin{figure*}[t!]
\centering
\includegraphics[width=18.4cm]{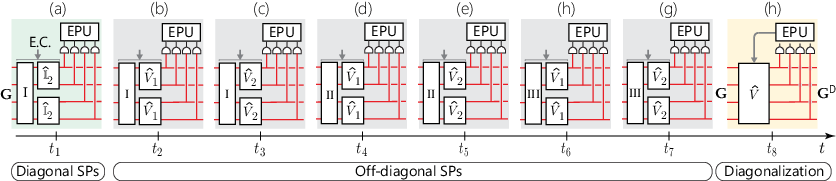}
\caption{Procedure for diagonalizing a $4\times4$ coherence matrix $\mathbf{G}$ via Stokes tomography, enumerated as 8~steps at time instances $\{t_{1},t_{2},\cdots,t_{8}\}$. The modal switch and subsequent unitaries are activated by an electrical control (E.C.) and the measured modal weights are stored in an electronic processing unit (EPU). (a) The modal switch is in setting~I and the modal weights are recorded to reconstruct the diagonal elements $G_{11},G_{22},G_{33}$, and $G_{44}$. (b) The modal switch is in setting~I, two unitaries $\hat{V}_{1}$ are implemented, and the modal weights are recorded to obtain the real parts of $G_{12}$ and $G_{34}$. (c) Same as (b), except that two unitaries $\hat{V}_{2}$ are implemented to obtain the imaginary parts of $G_{12}$ and $G_{34}$. (d,e) Same as (b,c) except that the modal switch is in setting~II to obtain $G_{13}$ and $G_{23}$. (f,g) Same as (b,c) except that the modal switch is in setting~III to obtain $G_{14}$ and $G_{23}$. (h) In a final step, the measurements from (a-g) stored in the EPU are used to calculate the unitary $\hat{V}$ that diagonalizes $\mathbf{G}$, which is implemented on chip to yield $\mathbf{G}^{\mathrm{D}}$.}
\label{fig:4x4DiagonalizationSequence}
\end{figure*}

The modal switch (itself a unitary operator) takes on one of three settings [Fig.~\ref{fig:4x4GeneralStokesRecon}(e)]:

\textit{Setting I:} The modes remain the same $\{|1\rangle,|2\rangle,|3\rangle,|4\rangle\}\rightarrow\{|1\rangle,|2\rangle,|3\rangle,|4\rangle\}$, and the modal switch is thus the identity operator. In this configuration, the pair of $2\times2$ unitaries $\hat{V}_{1}$ (or $\hat{V}_{2}$) are implemented on modes $\{|1\rangle,|2\rangle\}$ and $\{|3\rangle,|4\rangle\}$, thereby yielding the real (or imaginary) parts of the off-diagonal elements $G_{12}$ and $G_{34}$.

\textit{Setting II:} The modal switch implements the mapping $\{|1\rangle,|2\rangle,|3\rangle,|4\rangle\}\rightarrow\{|1\rangle,|3\rangle,|2\rangle,|4\rangle\}$; i.e., modes $|2\rangle$ and $|3\rangle$ are switched. The pairs of $2\times2$ unitaries are thus implemented on modes $\{|1\rangle,|3\rangle\}$ and $\{|2\rangle,|4\rangle\}$, thereby yielding the real and imaginary parts of the off-diagonal elements of $G_{13}$ and $G_{24}$.

\textit{Setting III:} The modal switch implements the mapping $\{|1\rangle,|2\rangle,|3\rangle,|4\rangle\}\rightarrow\{|1\rangle,|4\rangle,|2\rangle,|3\rangle\}$. The pairs of $2\times2$ unitaries are thus implemented on modes $\{|1\rangle,|4\rangle\}$ and $\{|2\rangle,|3\rangle\}$, thereby yielding the real and imaginary parts of the off-diagonal elements $G_{14}$ and $G_{23}$.

We are now in a position to describe the Stokes-tomographic diagonalization procedure in detail, as depicted in Fig.~\ref{fig:4x4DiagonalizationSequence}. In step~1 (at time $t_{1}$), we implement setting~I for the modal switch followed with no unitaries and record the modal weights. This step yields the diagonal elements of $\mathbf{G}$. In step~2 (at time~$t_{2}$) we retain setting~I for the modal switch and implement $\hat{V}_{1}$ on the first two modes ($|1\rangle$ and $2\rangle$ here) and $\hat{V}_{1}$ on the remaining two modes ($|3\rangle$ and $|4\rangle$), thereby obtaining the real parts of $G_{12}$ and $G_{34}$. In step~3 (at time~$t_{3}$), we retain the modal switch in setting~I but implement the unitaries $\hat{V}_{2}$, thus obtaining the imaginary parts of $G_{12}$ and $G_{34}$. In step~4 (at time~$t_{4}$), we employ setting~II for the modal switch, so that the first $\hat{V}_{1}$ is now implemented on modes $|1\rangle$ and $|3\rangle$, while the second $\hat{V}_{1}$ is implemented on modes $|2\rangle$ and $|4\rangle$, thus yielding the real parts of $G_{13}$ and $G_{24}$. Step~5 (at time~$t_{5}$) is the same as step~4 except that the unitaries $\hat{V}_{1}$ are replaced by $\hat{V}_{2}$, thereby yielding the imaginary parts of $G_{13}$ and $G_{24}$. In step~6 we employ setting~III for the modal switch, so that the first $\hat{V}_{1}$ is implemented on modes $|1\rangle$ and $|4\rangle$, while the second $\hat{V}_{1}$ is implemented on modes $|2\rangle$ and $|3\rangle$, thus yielding the real parts of $G_{14}$ and $G_{23}$. Finally, in step~7 we retain setting~III for the modal switch and replace $\hat{V}_{1}$ by $\hat{V}_{2}$, thereby obtaining the imaginary parts of $G_{14}$ and $G_{23}$.

We have thus reconstructed the coherence matrix $\mathbf{G}$ in only 7~steps. From the reconstructed coherence matrix $\mathbf{G}$ we calculate a final $4\times4$ unitary $\hat{V}$ (that which diagonalizes $\mathbf{G}$), and implement it in step~8 at $t_{8}$ to diagonalize $\mathbf{G}\rightarrow\hat{V}\mathbf{G}\hat{V}^{\dagger}=\mathbf{G}^{\mathrm{D}}$.

\subsection{Measurement results for four-mode diagonalization}

We have implemented the diagonalization procedure depicted in Fig.~\ref{fig:4x4DiagonalizationSequence} for the 8~coherence matrices $\mathbf{G}_{j}^{(k)}$, and we plot the measurement results in Fig.~\ref{fig:4ModeData}. For each coherence matrix, we plot separately the real and imaginary parts of the matrix elements, for both the reconstructed coherence matrices and their theoretically expected counterparts. For each coherence matrix, we implement the sequence of 7~steps needed for reconstruction via Stokes tomography (1~for the diagonal elements of $\mathbf{G}$, 6~for its off-diagonal elements). In Table~\ref{Table:4x4} we list the values of the rank, the reconstruction fidelity $\mathcal{F}$ (all of which are larger than $94\%$), theoretically expected field entropy $S_{\mathrm{th}}$, and the entropy extracted from the reconstructed coherence matrix $S_{\mathrm{meas}}$.

Once $\mathbf{G}_{j}^{(k)}$ is reconstructed, we compute the $4\times4$ unitary $\hat{V}$ that diagonalizes it, $\hat{V}\mathbf{G}_{j}^{(k)}\hat{V}^{\dagger}=\mathbf{G}_{j}^{\mathrm{D}}$. Although the computed diagonalizing unitary $\hat{V}$ is unique for a given $\mathbf{G}$, the decomposition of this $4\times4$ unitary $\hat{V}$ into a sequence of $2\times2$ unitaries is \textit{not} unique. We need to carry out this decomposition in order to implement $\hat{V}$ utilizing the MZI mesh.

In general, a $4\times4$ unitary can always be decomposed into a sequence of 6~unitaries \cite{Reck94PRL,Saleh25book}, each operating on only two modes, as depicted in Fig.~\ref{fig:UnitaryDecomposition}(a). The specific order of modes on which these $2\times2$ are implemented impacts the unitaries themselves. To simplify the implementation, we select the order of the $2\times2$ unitaries based on the order in which the modes are left in after implementing the unitaries $\hat{U}_{a}$ and $\hat{U}_{b}$. The modes $\{|1\rangle,|2\rangle,|3\rangle,|4\rangle\}$ after the unitary $\hat{U}_{a}$ are left in the order $\{|1\rangle,|3\rangle,|2\rangle,|4\rangle\}$ [Fig.~\ref{fig:FourModeSetup}(b)]. Therefore, we decompose the $4\times4$ unitary $\hat{V}$ for $\mathbf{G}_{j}^{(a)}$ ($j=1,\cdots,4$) first into $2\times2$ unitaries $\hat{V}_{13}$ and $\hat{V}_{24}$ to minimize the number of modal switches needed, followed by the unitaries $\hat{V}_{12}$ and $\hat{V}_{34}$, and finally the unitaries $\hat{V}_{14}$ and $\hat{V}_{23}$. We present the detailed decomposition for the example of $\mathbf{G}_{3}^{(a)}$ in the Appendix. We find that the final two unitaries $\hat{V}_{14}$ and $\hat{V}_{23}$ both correspond approximately to unitary operators and are thus not implemented [Fig.~\ref{fig:UnitaryDecomposition}(b)]. When we decompose the $4\times4$ unitary $\hat{V}$ for $\mathbf{G}_{j}^{(b)}$ ($j=1,\cdots,4$), we exploit the fact that the modes $\{|1\rangle,|2\rangle,|3\rangle,|4\rangle\}$ after the unitary $\hat{U}_{b}$ are left in the order $\{|1\rangle,|4\rangle,|2\rangle,|3\rangle\}$ [Fig.~\ref{fig:FourModeSetup}(c)]. Therefore, to reduce the number of modal switches we first decompose $\hat{V}$ into $2\times2$ unitaries $\hat{V}_{14}$ and $\hat{V}_{23}$, followed by the unitaries $\hat{V}_{13}$ and $\hat{V}_{24}$, and finally the unitaries $\hat{V}_{12}$ and $\hat{V}_{34}$. We present the detailed decomposition for the example of $\mathbf{G}_{3}^{(b)}$ in the Appendix, where we find that the final four unitaries $\hat{V}_{13}$, $\hat{V}_{24}$, $\hat{V}_{12}$, and $\hat{V}_{34}$ all correspond approximately to unitary operators and are thus not implemented [Fig.~\ref{fig:UnitaryDecomposition}(c)]. The results confirm that the Stokes-tomography-based procedure correctly diagonalizes the coherence matrices for four-mode light.

We plot in Fig.~\ref{fig:4ModeData} the modal weights (diagonal elements) of the reconstructed coherence matrix before and after its diagonalization, and compare the latter to the eigenvalues programmed into the field. We provide in Table~\ref{Table:4x4} the values of entropy $S_{\mathrm{D}}$ obtained from the diagonalized coherence matrices, which are in excellent agreement with the encoded values $S_{\mathrm{th}}$. 

\begin{table}[t!]
\caption{Rank, the reconstruction fidelity $\mathcal{F}$, the theoretically expected entropy $S_{\mathrm{th}}$, the entropy obtained from the reconstructed coherence matrices $S_{\mathrm{meas}}$, and the entropy obtained from the diagonalized coherence matrices $S_{\mathrm{D}}$.}\label{Table:4x4}
\setlength\extrarowheight{1pt}
\begin{tabular}
{|{c}||{c}|{c}|{c}|{c}|{c}|}
 \hline
  & rank & $\mathcal{F}$ & $S_{\mathrm{th}}$ & $S_{\mathrm{meas}}$ & $S_{\mathrm{D}}$ \\ \hline\hline

$\mathbf{G}_{1}^{(a)}$&1&0.94&0&0.33&0.117\\ \hline

$\mathbf{G}_{1}^{(b)}$&1&0.98&0&0.1&0.101\\ \hline

$\mathbf{G}_{2}^{(a)}$&2&0.98&0.88&0.994&0.919\\ \hline

$\mathbf{G}_{2}^{(b)}$&2&0.99&0.88&0.943&0.968\\ \hline

$\mathbf{G}_{3}^{(a)}$&3&0.99&1.485&1.504&1.565\\ \hline

$\mathbf{G}_{3}^{(b)}$&3&0.99&1.485&1.534&1.527\\ \hline

$\mathbf{G}_{4}^{(a)}$&4&0.99&1.846&1.854&1.868\\ \hline

$\mathbf{G}_{4}^{(b)}$&4&0.99&1.846&1.886&1.89\\ \hline

\end{tabular}\end{table}

\section{Discussion}

\subsection{Scaling of Stokes tomography with the number of modes}

Reconstructing $\mathbf{G}$ via Stokes tomography required $N_{\mathrm{m}}=3$~steps for $N=2$ and $N_{\mathrm{m}}=7$~steps for $N=4$. The question remains as to how the number of required measurements $N_{\mathrm{m}}$ scales generally with $N$. The scheme for $N=4$ in Fig.~\ref{fig:4x4GeneralStokesRecon} and Fig.~\ref{fig:4x4DiagonalizationSequence} provides a blueprint for estimating $N_{\mathrm{m}}$ more generally. Consider a field spanned by $N$~modes (assume first that $N$~is even). After the modal switch, the $N$~modes are divided into $\tfrac{1}{2}N$~pairs of modes. Each pair is associated with a particular off-diagonal element of $\mathbf{G}$, so that each setting of the modal switch provides access to $\tfrac{1}{2}N$~off-diagonal elements of $\mathbf{G}$. Each pair of modes after the modal switch is directed to a unitary $\hat{V}_{1}$, thereby yielding the real parts of $\tfrac{1}{2}N$~off-diagonal elements. Holding the same setting of the modal switch, we replace each $\hat{V}_{1}$ with $\hat{V}_{2}$ to obtain the imaginary parts of the same $\tfrac{1}{2}N$~off-diagonal elements of $\mathbf{G}$. In these two steps we acquire $2\times\tfrac{N}{2}=N$ modal SPs. Making use of $N-1$ settings of the modal switch that implement the appropriate permutations of the $N$~modes, we acquire all the $N(N-1)$ off-diagonal modal SPs, which requires $2\times(N-1)=2N-2$~steps. When $N=2$, we require only 1~setting of the input modes and and 2~steps to acquire the off-diagonal SPs. For $N=4$, we require 3~settings of the input modes, and thus acquire the off-diagonal SPs in 6~steps. Adding the step for acquiring the diagonal elements of $\mathbf{G}$, we have a total number of measurements $N_{\mathrm{m}}=2N-1$ to reconstruct $\mathbf{G}$. Adding the final step for diagonalization implies that $2N$~steps are needed to diagonalize an $N\times N$ coherence matrix. In other words, the number of required measurements for reconstructing an $N\times N$ coherence matrix is $\mathcal{O}(N)$ rather than $\mathcal{O}(N^{2})$ as commonly thought \cite{roques2024Light,Mor26arxiv}.
\begin{figure*}[p]
\centering
\includegraphics[width=17.6cm]{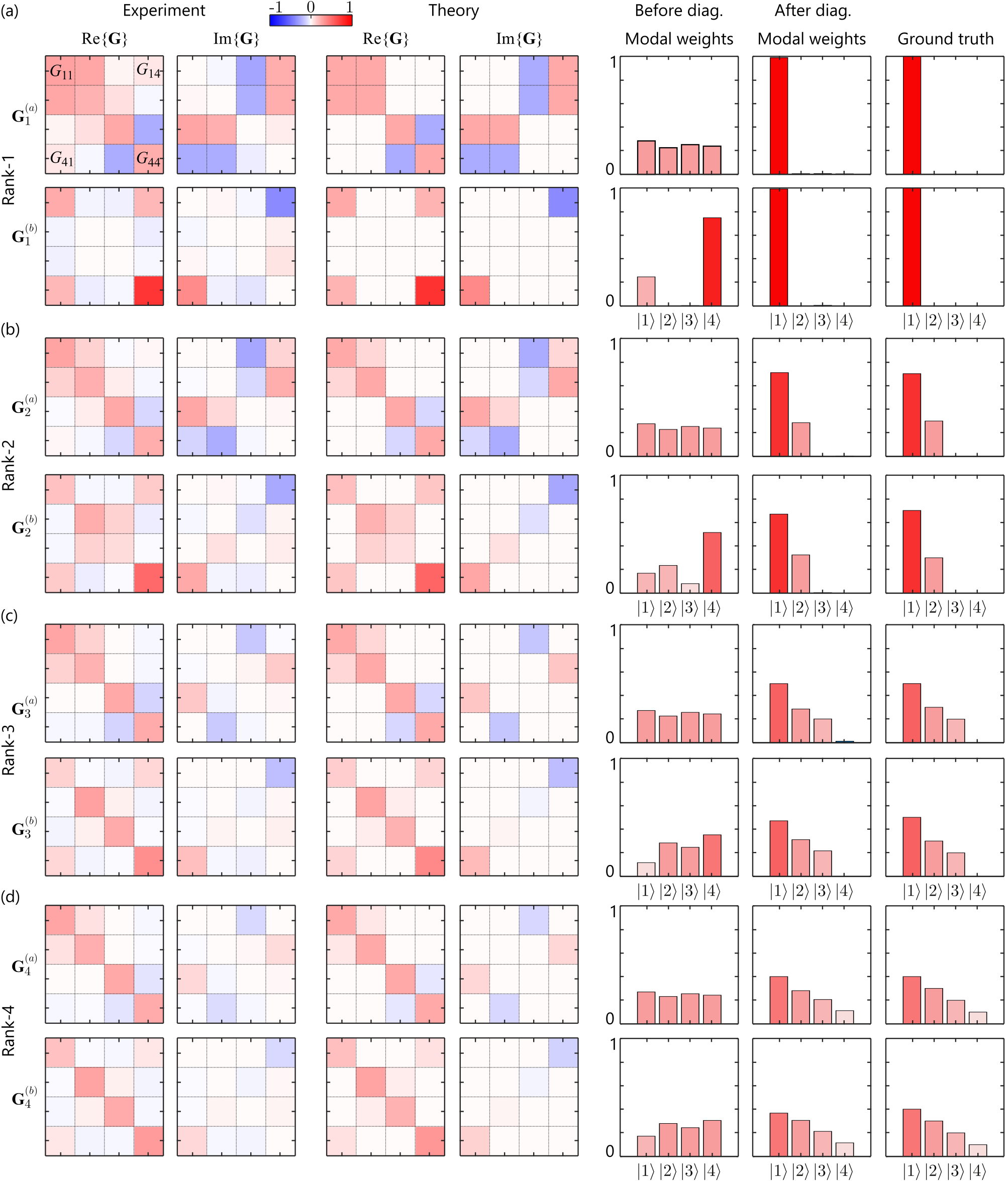} 
\caption{Stokes-tomography-based diagonalization of four-mode light. First and second columns are the real and imaginary parts of the reconstructed $4\times4$ coherence matrices, whereas the third and and fourth columns correspond to the theoretically expected values. The fifth column shows the modal weights of the reconstructed coherence matrices (the diagonal elements). The sixth column shows the modal weights of the diagonalized coherence matrices, and the seventh column that shows the theoretical eigenvalues of the synthesized fields. (a) Rank-1 fields $\mathbf{G}_{1}^{(a)}$ and $\mathbf{G}_{1}^{(b)}$ synthesized from $\mathbf{G}_{1}^{\mathrm{D}}$. (b) Rank-2 fields $\mathbf{G}_{2}^{(a)}$ and $\mathbf{G}_{2}^{(b)}$ synthesized from $\mathbf{G}_{2}^{\mathrm{D}}$. (c) Rank-3 fields $\mathbf{G}_{3}^{(a)}$ and $\mathbf{G}_{3}^{(b)}$ synthesized from $\mathbf{G}_{3}^{\mathrm{D}}$. (d) Rank-4 fields $\mathbf{G}_{4}^{(a)}$ and $\mathbf{G}_{4}^{(b)}$ synthesized from $\mathbf{G}_{4}^{\mathrm{D}}$.}
\label{fig:4ModeData}
\end{figure*}
Dealing with odd~$N$ requires one modification. After any setting of the modal switch, pairing the modes to unitaries $\hat{V}_{1}$ or $\hat{V}_{2}$ yields $\tfrac{1}{2}(N-1)$ pairs and leaves one unpaired mode that goes unutilized in that particular step. Consequently, each modal switch configuration yields $N-1$~off-diagonal modal SPs for odd~$N$ rather than $N$~off-diagonal modal SPs for even~$N$. We thus require $N$~settings for the modal switch, resulting in $2N$ steps for the off-diagonal SPs, and a total of $N_{\mathrm{m}}=2N+1$ to reconstruct $\mathbf{G}$ (and $2N+2$ to diagonalize $\mathbf{G}$).

\subsection{Future work}

A crucial task is now to compare the various approaches currently available for on-chip reconstruction of an unknown $N\times N$ coherence matrix $\mathbf{G}$. We have outlined here the number of steps required to diagonalize $\mathbf{G}$ when relying on Stokes tomography. Another recently proposed technique is `variational processing' \cite{roques2024Light,Mor26arxiv} that relies on a sequential search algorithm to estimate the eigenvalues of the coherence matrix. Such a strategy requires a much larger number of steps to diagonalize $\mathbf{G}$ compared to Stokes tomography. We recently demonstrated an alternative approach that we called single-shot Stokes tomography, realized for the case of $N=2$ \cite{Hashemi26OL}. In this scheme, rather than implementing a sequence of unitaries in time (as done here for example), the unitaries are implemented in parallel in a single step; that is, space is exchanged for time. Although this scheme obviously requires a larger chip area, it has the unique advantage of reconstructing the coherence matrix in a single step. The unitaries implemented on chip are fixed, and the reconstruction speed is limited only by the detector bandwidth.

An alternative approach that has \textit{not} been utilized so far is one inspired by the conventional scheme of double-slit interferometry. In double-slit interference, the field at the two slits is described by a $2\times2$ coherence matrix \cite{Gori06OL,Abouraddy17OE,Eberly17Optica,Abouraddy19Optica,Halder21OL}. The visibility of the double-slit interferogram (along with the modal weights) determine the magnitude of the off-diagonal element of the coherence matrix, while the shift in the central interference fringe (with respect to the diffraction pattern of a single slit) determines the phase of the off-diagonal element. The interferogram thus determines the complex off-diagonal element of $\mathbf{G}$. This scheme can also be implemented on chip with spatial modes confined to waveguides, where the interferogram is produced by a unitary such as $\hat{W}(\varphi)$ that superposes two modes and sweeps a relative phase between them. For a field spanned by $N$~modes, this interferogram is to be acquired for the $N(N-1)/2$ possible pairs of modes. To the best of our knowledge, this procedure has not been tested to date.

\begin{figure}[t!]
\centering
\includegraphics[width=8.5cm]{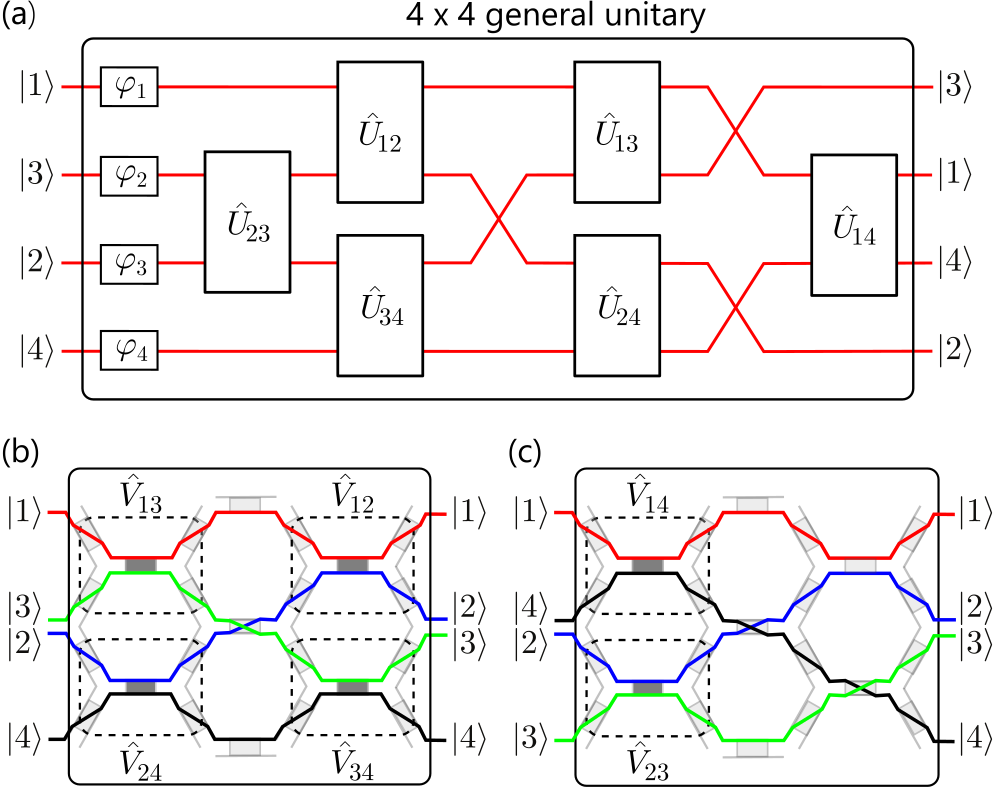} 
\caption{(a) A $4\times4$ general unitary on four-mode light can be decomposed into 4 phases $\varphi_{j}$ ($j=1,\cdots,4$) implemented on the modes $\{|1\rangle,|2\rangle,|3\rangle,|4\rangle\}$, followed by 6~unitaries, each is $2\times2$ and implemented on a pair of modes. We plot the sequence of $2\times2$ unitaries in a specific order, but other orders can also be selected as long as the 6~modal pairings are all involved. (b) A decomposition of the diagonalizing unitary $\hat{V}$ for the coherence matrices of the form $\mathbf{G}_{j}^{(a)}$ [Fig.~\ref{fig:FourModeSetup}(b)], depicted as a photonic circuit on the MZI mesh. The two remaining $2\times2$ unitaries $\hat{V}_{14}$ and $\hat{V}_{23}$ are approximately the identity operator (Appendix) and are thus not implemented. (c) Same as (b) for the coherence matrices $\mathbf{G}_{j}^{(b)}$. Here, the $2\times2$ unitaries $\hat{V}_{12}$, $\hat{V}_{13}$, $\hat{V}_{24}$, and $\hat{V}_{34}$ are all approximately the identity operator (Appendix), and are thus not implemented.}
\label{fig:UnitaryDecomposition}
\end{figure}

\subsection{Conclusion}

In conclusion, we have demonstrated that Stokes tomography allows for the on-chip diagonalization of a multimode partially coherent field. Starting with an arbitrary partially coherent field spanned by $N$ modes and described by an $N\times N$ coherence matrix $\mathbf{G}$, diagonalization of $\mathbf{G}$ takes $N_{\mathrm{m}}+1$ steps. Diagonalization takes place in two stages. In the first stage, the coherence matrix is reconstructed in $N_{\mathrm{m}}$ steps, each step involving implementing a unitary from a predetermined sequence followed by a measurement of the modal weights. At the end of this first stage, the measurements are combined to reconstruct $\mathbf{G}$. In the second stage (which occupies a single step), an $N\times N$ unitary $\hat{V}$ is computed from the reconstructed $\mathbf{G}$ and implemented on chip. When $N$~is even, $N_{\mathrm{m}}=2N-1$ steps are required to reconstruct $\mathbf{G}$, and when $N$~is odd, we have $N_{\mathrm{m}}=2N+1$. Therefore, diagonalization occurs in $2N$~steps when~$N$ is even, and in $2N+2$ steps when~$N$ is odd. We have implemented this Stokes-tomographic diagonalization scheme with two-mode ($N=2$) and four-mode ($N=4$) light, while varying the rank, entropy, and structure of $\mathbf{G}$. Additionally, we have confirmed the diagonalization interferometrically and verified that the field is left in the coherent-mode representation -- all while retaining a copy of the initial field $\mathbf{G}$ available for separate on-chip processing. The on-chip platform provides interferometric stability and scalability to optical fields spanned with a large number of modes $N$. These results usher in a new era for on-chip utilization of partially coherent light by establishing Stokes tomography as a versatile tool for on-chip processing of multimode partially coherent light for applications in optical communications \cite{Nardi22OL,Harling25APLP}, cryptography \cite{Liu25LPR}, computing \cite{Dong24Nature}, and spectroscopy \cite{Miller25Optica}.

\section*{Appendix: Decomposition of the diagonalizing unitary for four-mode light}

To elucidate how the unitary $\hat{V}$ is implemented on chip, we provide here a specific example from our measurements. Consider the coherence matrix $\mathbf{G}_{3}^{(a)}$ resulting from implementing the unitary $\hat{U}_{a}$ on the diagonal coherence matrix $\mathbf{G}_{3}=\mathrm{diag}\{0.5,0.3,0.2,0\}$. After reconstruction of $\mathbf{G}_{3}^{(a)}$, we calculated the unitary $\hat{V}$ that diagonalizes it, which is given by:
\begin{equation}
\left(\!\!\!\begin{array}{cccc}
-0.05-i0.6&-0.05-i0.42&-0.52-i0.02&0.44\\
-0.06+i0.41&-0.17-i0.53&0.44-i0.17&0.55\\
-0.05-i0.51&-0.25-i0.38&0.5-i0.15&-0.51\\
-0.06+i0.5&0.03-i0.56&-0.48-i0.05&-0.5
\end{array}\!\!\!\!\right),
\end{equation}
where we list all the real and imaginary parts of the elements of $\hat{V}$ only to second decimal place. While $\hat{V}$ is unique, its decomposition into a sequence of $2\times2$ unitaries is \textit{not}.

We decompose $\hat{V}$ into a sequence of $2\times2$ unitaries by first finding the two $2\times2$ unitaries $\hat{V}_{13}$ and $\hat{V}_{24}$ that eliminate the off-diagonal elements $V_{13}$ and $V_{24}$ from $\hat{V}$, which are given by: 
\begin{equation}
\hat{V}_{13}=\left(\begin{array}{cc}
0.75&0.08-i0.65\\
0.08+i0.65&-0.75
\end{array}\right),
\end{equation}
\begin{equation}
\hat{V}_{24}=\left(\begin{array}{cc}
0.71&-0.21+i0.67\\
-0.21-i0.67&-0.71
\end{array}\right).
\end{equation}
Once again, we list all the real and imaginary parts of the elements of $\hat{V}$ only to second decimal place. Next, we find the two $2\times2$ unitaries $\hat{V}_{12}$ and $\hat{V}_{34}$ that eliminate the off-diagonal elements $V_{12}$ and $V_{34}$ from $\hat{V}$, which are given by:
\begin{equation}
\hat{V}_{12}=\left(\begin{array}{cc}
0.8&0.6-i0.08\\
0.6+i0.08&-0.8
\end{array}\right),
\end{equation}
\begin{equation}
\hat{V}_{34}=\left(\begin{array}{cc}
0.73&-0.68\\
-0.68&-0.73
\end{array}\right).
\end{equation}
Finally, we find the two $2\times2$ unitaries $\hat{V}_{14}$ and $\hat{V}_{23}$ that eliminate the off-diagonal elements $V_{14}$ and $V_{23}$ from $\hat{V}$, which are given by:
\begin{equation}
\hat{V}_{14}=\left(\begin{array}{cc}
0.99&0.04+i0.02\\
0.04-i0.02&-0.99
\end{array}\right),
\end{equation}
\begin{equation}
\hat{V}_{23}=\left(\begin{array}{cc}
0.99&0.12+i0.1\\
0.12-i0.1&-0.99
\end{array}\right).
\end{equation}

Note that $\hat{V}_{14}$ and $V_{23}$ are close to identity matrices. In constructing a $4\times4$ unitary $\hat{V}$, we assemble the first 4 unitaries: (1) implement the $2\times2$ unitary $\hat{V}_{13}$ on modes $(1,3)$ and the $2\times2$ unitary $\hat{V}_{24}$ on modes $(2,4)$; (2) switch modes $(2,3)$; and (3) implement the $2\times2$ unitary $\hat{V}_{12}$ on modes $(1,2)$ and the $2\times2$ unitary $\hat{V}_{34}$ on modes $(3,4)$. We do not implement the final two unitaries $\hat{V}_{14}$ and $\hat{V}_{23}$. This construction yielded the results plotted in Fig.~\ref{fig:4ModeData} for $\mathbf{G}_{j}^{(a)}$; see Fig.~\ref{fig:UnitaryDecomposition}(b). 

We obtain from the coherence matrix $\mathbf{G}_{3}^{(b)}$ the diagonalizing matrix $\hat{V}$: 
\begin{equation}
\left(\!\!\begin{array}{cccc}
0.23+i0.43&-0.12-i0.1&-0.07-i0.13&0.85\\
-0.05-i0.09&-0.44-i0.74&-0.45-i0.18&-0.15\\
0.06+i0.02&0.25-i0.43&0.09+i0.96&0.09\\
0.42+i0.76&-0.02-i0.02&0.03&-0.5
\end{array}\!\!\right).
\end{equation}
We decompose this unitary into a sequence of 6 unitaries $\hat{V}_{14}$, $\hat{V}_{23}$, $\hat{V}_{13}$, $\hat{V}_{24}$, $\hat{V}_{12}$, and $\hat{V}_{34}$, all $2\times2$, given by:
\begin{equation}
\hat{V}_{14}=\left(\begin{array}{cc}
0.5&0.41-i0.76\\0.41+i0.76&-0.5
\end{array}\right),
\end{equation}

\begin{equation}
\hat{V}_{23}=\left(\begin{array}{cc}
0.87&0.39-i0.3\\0.39+i0.3&-0.87
\end{array}\right),
\end{equation}

\begin{equation}
\hat{V}_{13}=\left(\begin{array}{cc}
0.99&0.09+i0.06\\0.09-i0.06&-0.99
\end{array}\right),
\end{equation}

\begin{equation}
\hat{V}_{24}=\left(\begin{array}{cc}
0.99&-i0.02\\i0.02&-0.99
\end{array}\right),
\end{equation}

\begin{equation}
\hat{V}_{12}=\left(\begin{array}{cc}
0.98&-0.18+i0.01\\-0.18-i0.01
\end{array}\right),
\end{equation}

\begin{equation}
\hat{V}_{34}=\left(\begin{array}{cc}
0.99&0.04\\0.04&-0.99
\end{array}\right).
\end{equation}
The last 4~unitaries are approximately identity matrices, and are thus not implemented in the decomposition of $\hat{V}$; see Fig.~\ref{fig:UnitaryDecomposition}(c).

A.S. thanks Alireza Fardoost and Fatemeh Ghaedi Vanani for fruitful conversations.

U.S. Office of Naval Research (ONR) N00014-20-1-2789.

The authors declare no conflicts of interest.

Data underlying the results presented in this paper are not publicly available at this time but may be obtained from the authors upon reasonable request.

\bibliography{diffraction}

\end{document}